\documentclass[reprint,bibnotes,amsmath,amssymb,prd,nofootinbib]{revtex4-2}
\usepackage{amsmath,amssymb,graphics,epsfig,subfigure}
\usepackage[table,xcdraw]{xcolor}
\usepackage{color}
\usepackage{hyperref}

\begin{document}

\thispagestyle{empty}

\begin{center}

\title{Dynamic behaviours of black hole  phase transitions near   quadruple points}

\date{\today}
\author{Jiayue Yang$^{1,2}$ \footnote{E-mail: j43yang@uwaterloo.ca},
       and Robert B. Mann$^{1,2}$ \footnote{E-mail: rbmann@uwaterloo.ca}}

 \affiliation{
$^{1}$Department of Physics and Astronomy, University of Waterloo, Waterloo, Ontario, Canada, N2L 3G1\\
$^{2}$
Perimeter Institute, 31 Caroline Street North, Waterloo, ON, N2L 2Y5, Canada}

\begin{abstract}
Treating the horizon radius as an order parameter in a thermal fluctuation, the free energy landscape model sheds light on the 
dynamic behaviour of black hole phase transitions. Here we carry out the first investigation of the dynamics
of the recently discovered multicriticality in black holes.
We specifically consider  black hole quadruple points   in  
$D=4$ Einstein gravity 
coupled to non-linear electrodynamics. We observe thermodynamic phase transitions between the four stable phases at a quadruple point as well as the weak and strong oscillatory phenomena by numerically solving the Smoluchowski equation describing the evolution of the probability distribution function.   We analyze the dynamic evolution of the different phases at various ensemble temperatures and find that the probability distribution of a final stationary state is closely tied to the structure of its off-shell Gibbs free energy. 
\end{abstract}

\pacs{?}

\maketitle
\end{center}

\section{Introduction}

Over the past five decades, noteworthy progress has been made in black hole thermodynamics after the four laws of black hole thermodynamics were proposed 
 \cite{Hawking:1974rv,Bekenstein:1972tm,Hawking,Bekenstein,Bardeen}. A particularly interesting subfield is that of critical phenomena in black hole thermodynamics. The Hawking-Page phase transition 
 \cite{Hawking:1982dh}, which describes the transition between thermal radiation and large black holes resembling the solid-liquid phase transition, was a seminal investigation into this topic.  
 In more recent years, the study of critical phenomena has become much better understood in the context of 
 black hole chemistry 
 \cite{Kubiznak:2016qmn}, whose main idea is to treat the black hole mass as enthalpy \cite{Kastor}, and the  cosmological constant and its conjugate variable 
 as the respective 
 thermodynamic pressure and volume of the black hole system. Under this interpretation, there is a phase transition between large and small charged anti-de
Sitter (AdS) black holes, which coincides with the gas-liquid phase transition in the Van der Waals model \cite{Kubiznak,Cai,Chamblin:1999tk,Chamblin:1999hg,Wu:2000id,Dolan:2011xt}.

Quite recently the underlying kinetics of black hole phase transitions have been studied
based on the free energy landscape model \cite{Li,LiWang,Li:2020spm,Li:2021vdp}. In this approach,  black hole phase transitions can be understood to take place due to thermal fluctuations,  with the black hole horizon radius regarded as the order parameter formulating the free energy
landscape.  The dynamic behaviour of black hole phase transitions can be understood by solving the Smoluchowski equation \cite{Wei:2020rcd,Yang:2021ljn,Taniguchi,Erickcek}, which is essentially the probabilistic Fokker-Planck equation depicting the diffusion process of a system given some potential barriers. For black hole phase transitions,  the effective potential barrier is described by the off-shell Gibbs free energy,  which is defined as a continuous function of the black hole horizon radius at some ensemble temperature.

The discovery that black holes can have triple points has further supported the motivation for treating black holes as chemical systems  \cite{Altamirano,Wei2,Frassino:2014pha}.  Black hole triple points are where three stable phases -- small, intermediate, and large horizon size --   merge together, resembling the triple point of water where solid, liquid, and gas coexist at a particular pressure and temperature. The 
dynamic behaviour of black hole triple points was recently studied \cite{Wei:2021bwy},  where initially small, intermediate, or large black holes were found to be able to 
transit to the other two coexistent phases at the triple point, indicating that thermodynamic phase transitions can indeed occur dynamically.
Both weak and strong oscillatory behaviour were observed. 
In the former (weak)  case,  the probability of a non-initial state becomes maximal before decaying to stationarity with its probability never exceeding that of the initial state, whereas in the latter (strong) case the probability of a non-initial state exceeds that of the initial state.

Multicrticality is the most recent discovery to emerge from black hole chemistry. Multicritical points occur when several distinct phases in a system merge at a particular set of thermodynamic parameters, generalizing the notion of a triple point.  A variety of settings, including 4-dimensional Einstein gravity coupled to non-linear electrodynamics \cite{Tavakoli:2022kmo},   multiply rotating Kerr-AdS black holes \cite{Wu:2022bdk}, and   spherically symmetric black holes in Lovelock gravity \cite{Wu:2022plw}
have all been shown to have black hole solutions exhibiting multicritical behaviour.
Such behaviour in black hole physics is much less understood than tricriticality and other black hole phase behaviour, and warrants further study.


Motivated by the above, we investigate the dynamic behaviour of black hole quadruple points in the context of 4-dimensional Einstein gravitational theory coupled to non-linear electrodynamics, potentially providing some insights for our understanding of the nature of black hole phase transitions.  
We find that the early evolution of the probability distribution of black hole phases near the quadruple point depends on the first passage time. This we find is related to the size of the Gibbs potential barrier between phases. We also find that the distribution of the final stationary phases is determined by the 
relative heights of the potential wells of the off-shell Gibbs free energy.

\section{Quadruple Points in Black Hole Phase Transitions}

We consider non-linear electromagnetic fields in 4-dimensional Einstein gravity. The action is given by  \cite{Gao:2021kvr} 
\begin{align}\label{act1}
 S
 =&\int d^4x[\sqrt{-g}(R-2\Lambda-\sum^N_{i=1} \alpha_i(F^2)^i)]
\end{align}
where $\alpha_i$ are dimensional coupling constants, $F^2=F_{\mu\nu}F^{\mu\nu}$, with the electromagnetic field tensor $F_{\mu\nu}\equiv \partial_\mu A_\nu-\partial_\nu A_\mu$, and $A_\mu$ is the electromagnetic four-potential. The sum may truncate at some finite value of $N$, but it is also possible to consider $N\to\infty$ (for example, as in Born-Infeld electrodynamics). 
The coupling constants 
$\alpha_i$ can be regarded as independent thermodynamic variables, each with their own conjugates.
The existence of multicritical points depends on the relative choices of the $\alpha_i$ in addition to the other thermodynamic variables.

We consider the following ansatz describing spherically symmetric static black holes 
\begin{align}
    ds^2=-U(r)dt^2+\frac{1}{U(r)}dr^2+r^2d\Omega^2
\end{align}
\begin{align}
    A_\mu=[\Phi(r), 0, 0, 0]
\end{align}
If the spacetime is asymptotically AdS, and  $\Phi$   asymptotically vanishes, we can assume
\begin{align}
\label{eq12_2}
    \Phi=\sum_{r=1}^K b_ir^{-i}\qquad U=1+\sum_{i=1}^K c_i r^{-i}+\frac{r^2}{l^2}
\end{align}
 It is then possible to solve the field equations for the coefficients $b_i$ and $c_i$ in terms of the mass $M$, charge $Q$, and the coupling constants $\alpha_i$. 
Any given solution depends on a finite number of parameters that can be indexed either by $N$ 
(the number of $\alpha_i$ couplings in the action) or by $K$
(the number of parameters in the 
solution), along with $M$ and $Q$. In the latter case the expressions for
the metric and gauge field are explicitly given by \eqref{eq12_2} whereas in the former case these expressions are determined implicitly.
In either case, multicritical thermodynamic phenomena can be obtained by making 
appropriate choices of a finite number of $\alpha_i$, with the values of $K$ or $N$
determining the maximal degree of multicriticality
\cite{Tavakoli:2022kmo}.

In the following context, we focus on a setting in which the $\alpha_i$ are nonzero and independent for $i\leq 7$; for $i\geq 8$ all remaining $\alpha_i$ are determined in terms of these quantities. This  case  is known to admit quadruple points \cite{Tavakoli:2022kmo}.  This provides us with the simplest non-trivial example of multicritical dynamical  behaviour.
The solution is
\begin{align}
\Phi = & \frac{Q}{r} + \frac{b_5}{r^5} + \frac{b_9}{r^9} + \frac{b_{13}}{r^{13}} + \frac{b_{17}}{r^{17}} + \frac{b_{21}}{r^{21}} + \frac{b_{25}}{r^{25}} \\
    U=&1-\frac{2M}{r}+\frac{Q^2}{r^2}+\frac{b_5 Q}{2r^6}+\frac{b_9Q}{3r^{10}}+\frac{b_{13}Q}{4r^{14}}\nonumber \\
    &+\frac{b_{17}Q}{5r^{18}}+\frac{b_{21}Q}{6r^{22}}+\frac{b_{25}Q}{7r^{26}}+\frac{r^2}{l^2}
\end{align}
where the $b_i$ are known in terms of the $\alpha_i$ and $Q$ \cite{Tavakoli:2022kmo}, and 
$b_i = 0$ for $i>25$.

The thermodynamic temperature $T$, entropy $S$, pressure $P$, and volume $V$ are given by 
\begin{align}
        T=&\frac{1}{4\pi r_+}( 1+\frac{3r_+^2}{l^2}-\frac{Q^2}{r_+^2}-5\frac{b_5 Q}{2r_+^6}-3\frac{b_9Q}{r_+^{10}}\nonumber \\
        &-13\frac{b_{13}Q}{4r_+^{14}}-17\frac{b_{17}Q}{5r_+^{18}}-7\frac{b_{21}Q}{2r_+^{22}}-25\frac{b_{25}Q}{7r_+^{26}})\\
        S=&\pi r_+^2,\quad\quad P=\frac{3}{8\pi l^2},\quad\quad V=\frac{4\pi r_+^3}{3}
\end{align}
and the equation of state reads
 \begin{align}
    P=&\frac{T}{2r_+}-\frac{1}{8\pi r_+^2}+\frac{Q^2}{8\pi r_+^4}+\frac{5b_5Q}{16\pi r_+^8}+\frac{3b_9Q}{8\pi r_+^{12}}\nonumber \\
    &+\frac{13b_{13}Q}{32\pi r_+^{16}}+\frac{17b_{17}Q}{40\pi r_+^{20}}+\frac{7b_{21}Q}{16\pi r_+^{24}}+\frac{25b_{25}Q}{56\pi r_+^{28}}
\end{align}
where all formulae are written in  Planck
units, $Q$ is the black hole charge, $r_+$ is the horizon radius, the size of the black hole.

\begin{figure}
\includegraphics[width=7.0cm]{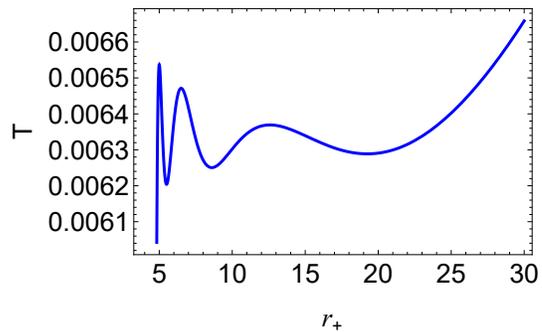}
\caption{Black hole quadruple points with parameters  $P=0.0000689999421$, $Q=6.751117513$ in the $T$-$r_+$ diagram. \label{tr}}
\end{figure}

The Gibbs free energy $G=M-TS$,
whose global minimum as a function of $T$ 
 determines the thermodynamically stable state,
provides a diagnostic for finding multicritical behaviour \cite{Tavakoli:2022kmo, Wu:2022bdk}. 
   The presence of $(N-1)$ swallowtails in the $G-T$ diagram indicates $N$ distinct phases, with 
the self-intersection points of the swallowtails signifying    
first order phase transitions
 between two distinct phases. Since each cusp in a swallowtail corresponds to an extremum of $T$ as a function of $r_+$, $N$ distinct phases 
require $2N-2$ distinct extrema in $T(r_+)$. 
  If two adjacent inflections of $T(r_+)$ occur at the same temperature, then the intersection points of corresponding swallowtails will merge, yielding a tricritical point.  If 
 this occurs for  
  $j \leq N$ different 
inflections, the $j$ 
  swallowtails will have a common self-intersection, corresponding to a $j$-th order multicritical point. 

These critical points can be found by finely tuning the thermodynamic parameters. We can obtain a quadruple point, where the $T (r_ +)$ curve has three pairs of local
maxima and minima equivalently, shown in Fig.\ref{tr}, by specifying 
\begin{align}
    \begin{cases}
    P=0.0000689999421\\
        Q = 6.751117513\\
        b_5 = -5078.980603\\
b_9 = 
 6.054804813\times10^6\\
        b_{13} = -4.131510152\times10^9\\
        b_{17} = 
 1.399821234\times10^{12}\\b_{21} = -2.014970449\times 10^{14}\\
 b_{25} \
= 1.016449472\times 10^{16}
    \end{cases}
\label{values}    
\end{align}

To understand the dynamics of quadruple points, we consider a canonical ensemble at some fixed temperature $T_{\rm E}$, which is not required to be the Hawking temperature $T$. Under the values of the parameters we choose (\ref{values}), the temperature range in which quadruple points can exist is $T_{\rm E}\in (0.00628883, 0.00636918 )$, as shown in Fig.~\ref{tr}.  Regarding the horizon radius   as the order parameter, we  generalize the on-shell Gibbs free energy to include black hole spacetimes with all radii. The off-shell Gibbs free energy $ G_{\rm L}$ is then defined in terms of the ensemble temperature $T_{\rm E}$ instead of the Hawking temperature $T$, obtaining
\begin{align}
 G_{\rm L} &= M-T_{\rm E}S  \label{glll}  \nonumber\\
&= \frac{1}{840 r_+^{25}}(60 b_{25} Q + 70 b_{21} Q r_+^4 + 84 b_{17} Q r_+^8 \nonumber \\
&+  105 b_{13} Q r^{12}
+ 140 b_9 Q r_+^{16} + 210 b_5 Q r_+^{20}\nonumber\\ & + 420 Q^2 r_+^{24}
    + 420 r_+^{26} 
    + 1120 P \pi r_+^{28}) 
    - T_{\rm E} \pi r_+^2 
\end{align}
 where $M$ is the black hole mass  
(the
conserved charge associated with the timelike Killing vector $\xi = \partial_t$ \cite{Tavakoli:2022kmo}), and $S$ is the entropy of black hole. We see from  (\ref{glll}) that the off-shell Gibbs free energy, as a function of $r_+$, also depends on the black hole charge $Q$, thermodynamic pressure $P$, and the ensemble temperature $T_{\rm E}$.
\begin{widetext}
\begin{center}
\begin{figure}[!h]
\center{
\subfigure[]{\label{gb}
\includegraphics[width=7.0cm]{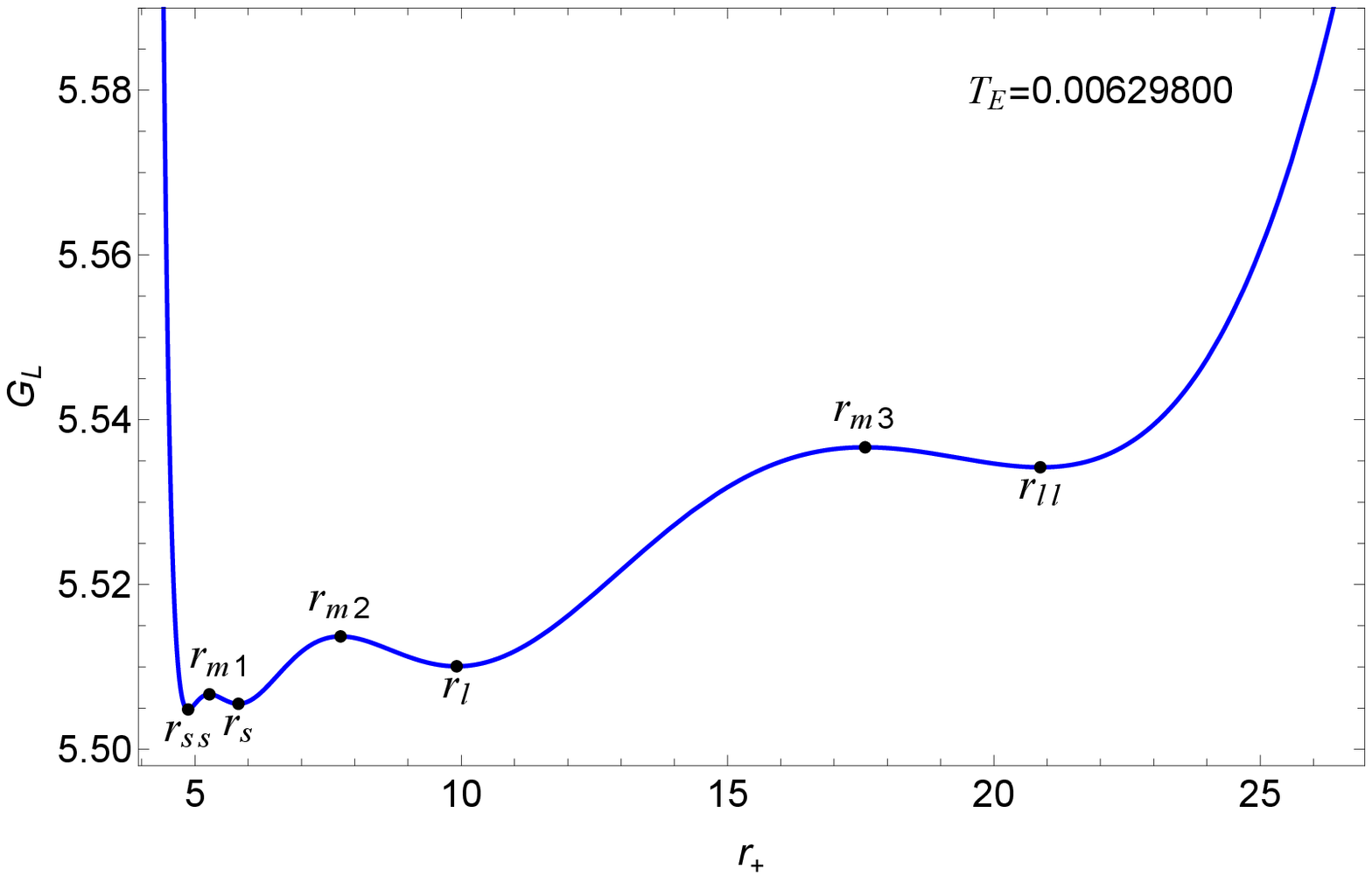}}
\subfigure[]{\label{ga}
\includegraphics[width=7.0cm]{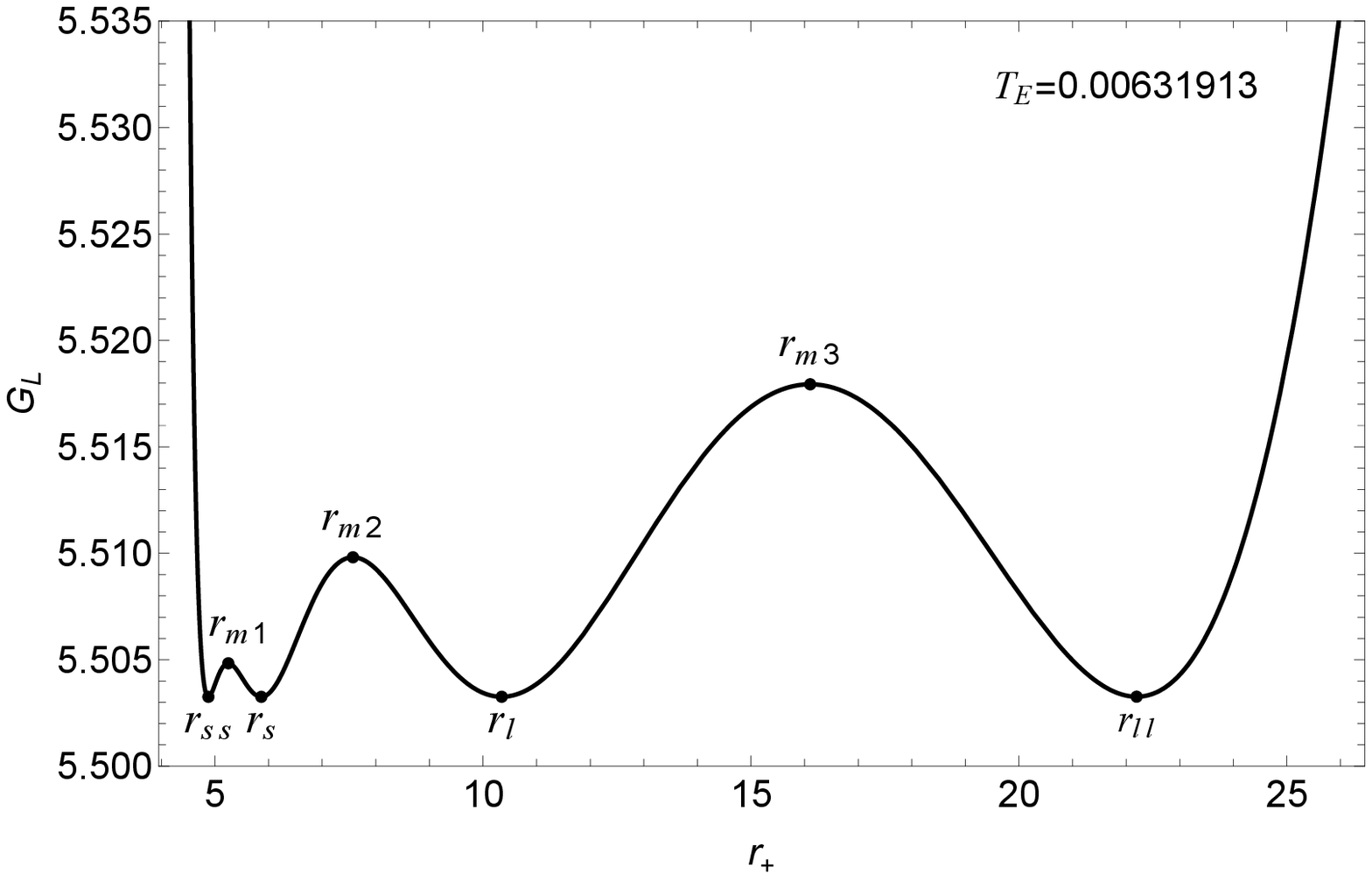}}\\
\subfigure[]{\label{gc}
\includegraphics[width=7.0cm]{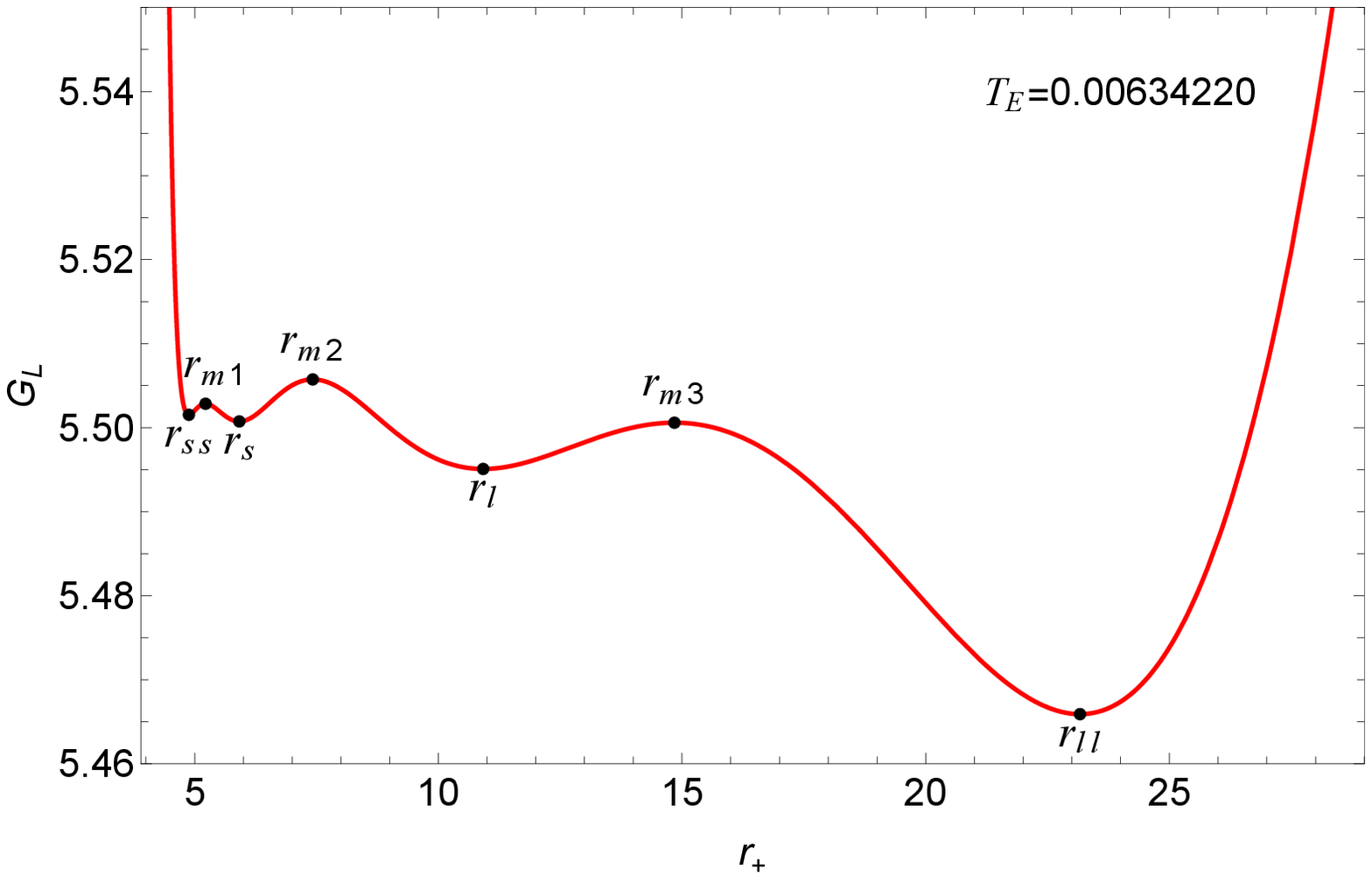}}
\subfigure[]{\label{gd}
\includegraphics[width=7.0cm]{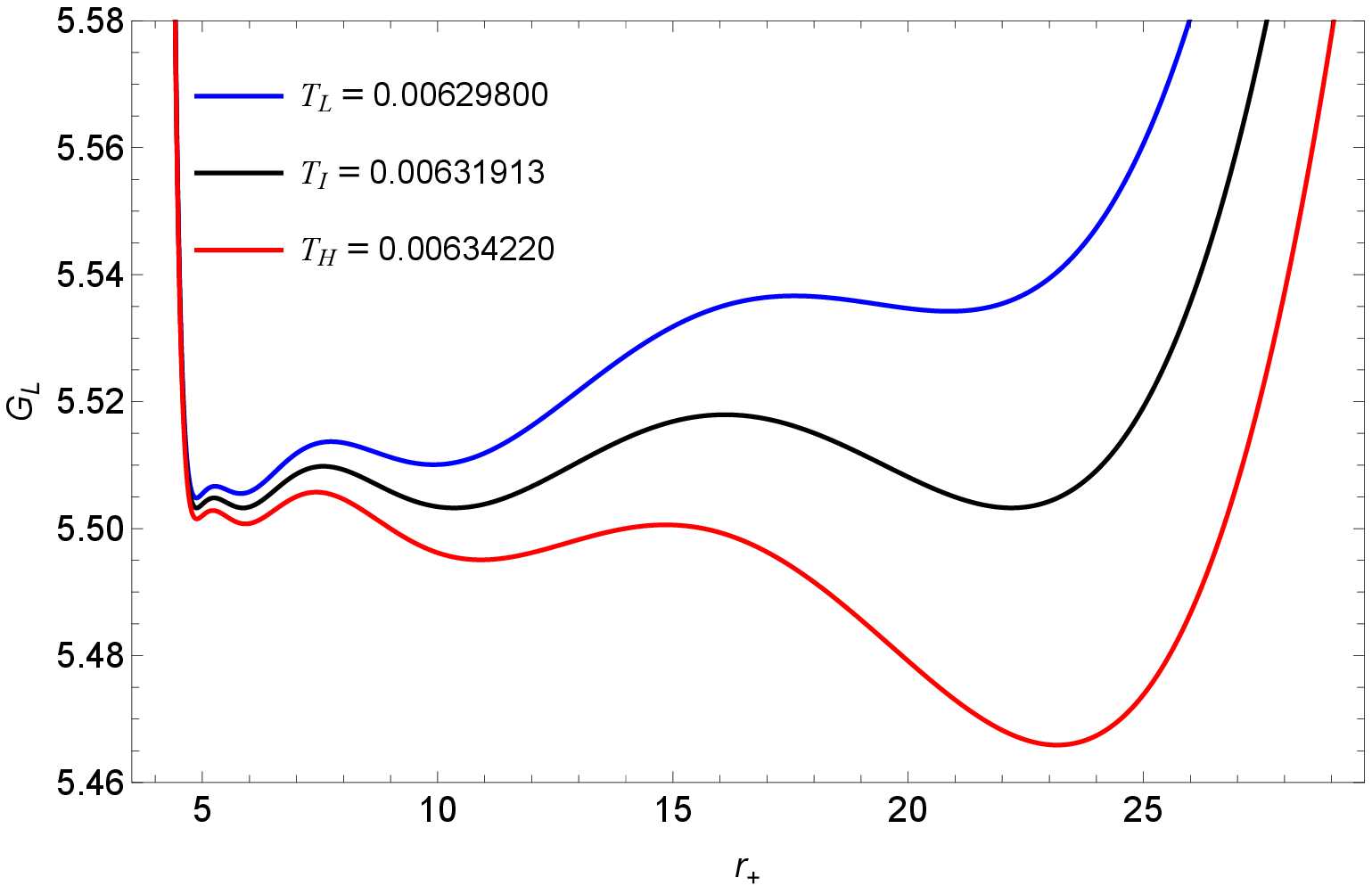}}}
 \caption{Off-shell Gibbs free energy at ensemble temperatures (a) $T_{\rm E}=0.006298$, (b) $T_{\rm E}=0.00631913$,  (c) $T_{\rm E}=0.0063422$. (d) Off-shell Gibbs free energy at three different ensemble temperatures are presented in one panel with low temperature $T_{\rm L}=0.006298$, intermediate temperature $T_{\rm I}=0.00631913$, and high temperature $T_{\rm H}=0.0063422$. }
\label{Gibbs}
\end{figure}
\end{center}
\end{widetext}

A union of black hole spacetimes with a range of arbitrary horizon  
radii constitutes the so-called black hole landscape.  On this landscape,
the off-shell free energy is interpreted as the effective potential in the black hole phase transition process, and the horizon radius is regarded as the order parameter. We depict the off-shell free energy $G_{\rm L}$ v.s. the horizon radius $r_+$  at a quadruple point with $P=0.0000689999421$, $Q=6.751117513$ (as given by \eqref{values}) 
at various ensemble temperatures, as shown in Fig.~\ref{Gibbs}.
Only the local extrema on the $G_{\rm L}$ curve correspond to physical black holes, which are on-shell solutions to the field equations 
that follow from \eqref{act1}. On each curve, there are four local minima,  i.e. $r_{ {k} }$=$r_{ ss}$, $r_{ s}$, $r_{ l}$, and $r_{ ll}$, representing four thermodynamic local stable black holes in order of increasing horizon size,  and three local maxima, i.e. $r_{ m1}$, $r_{ m2} $ and $r_{ m3}$, representing three unstable black hole phases or what we call intermediate transition states. Other points on the curve denote off-shell solutions,   which are transient black hole states during the phase transition process.

\begin{widetext}

\begin{center}
\begin{figure}
\center{\subfigure[]{\label{tmill}
\includegraphics[width=7.0cm]{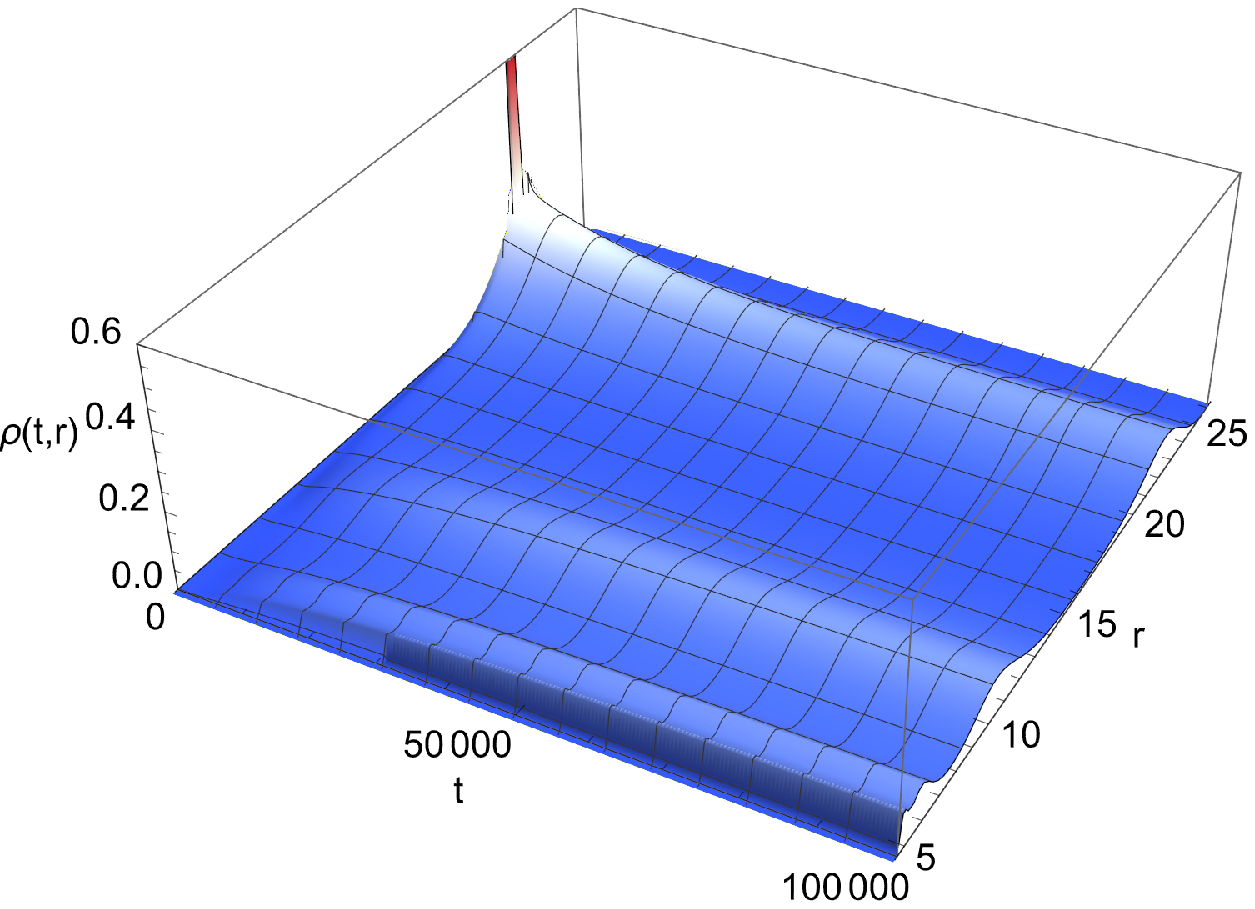}}
\subfigure[]{\label{tmil}
\includegraphics[width=7.0cm]{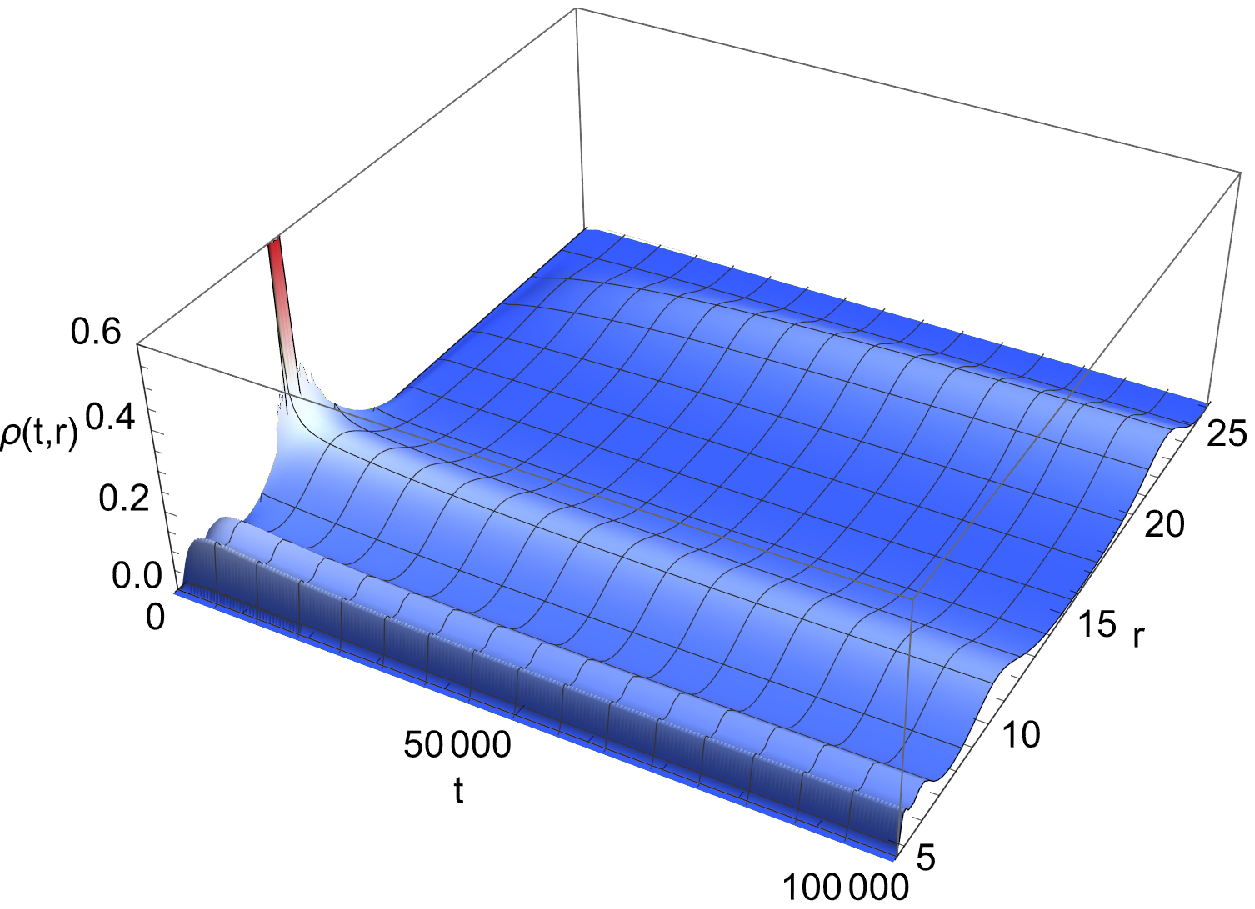}}\\
\subfigure[]{\label{tmis}
\includegraphics[width=7.0cm]{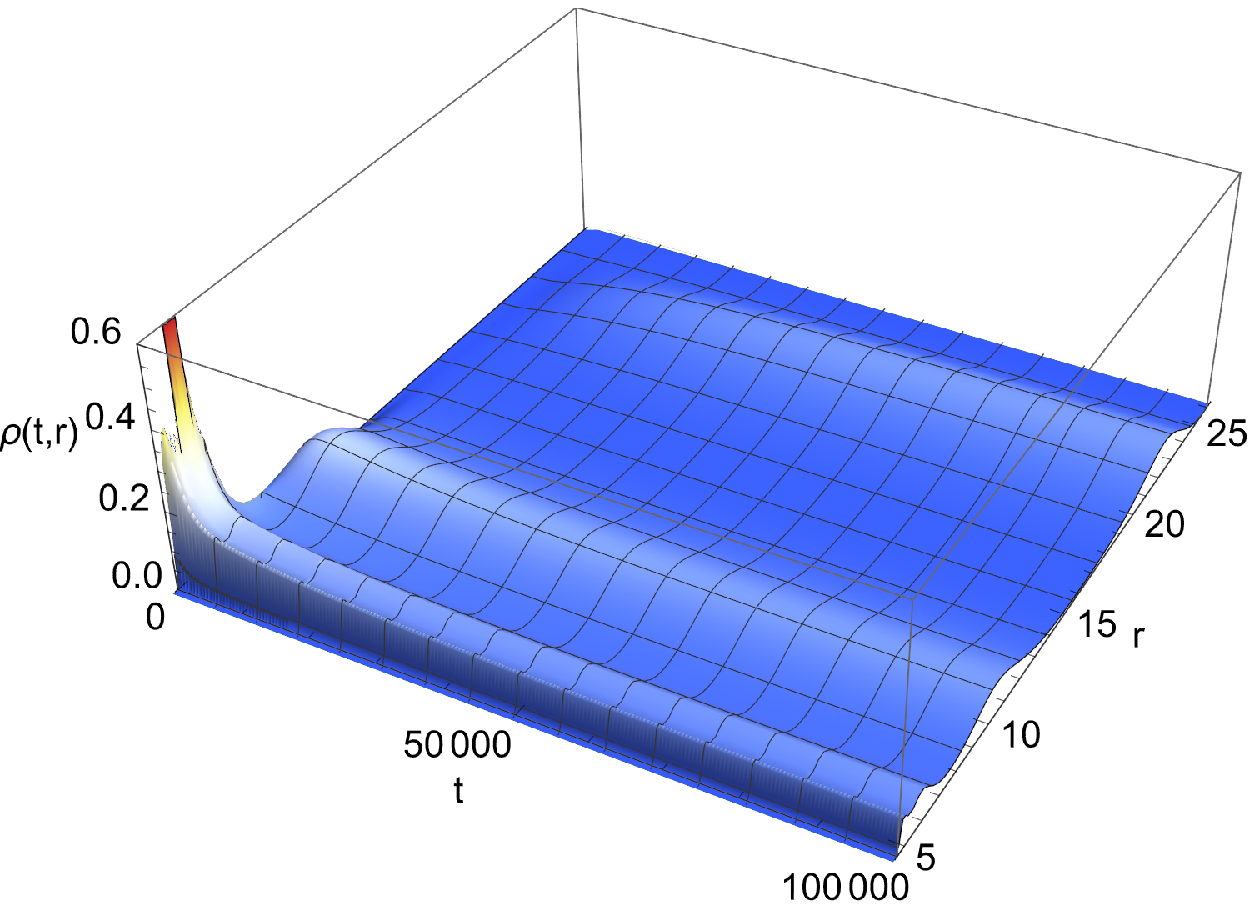}}
\subfigure[]{\label{tmiss}
\includegraphics[width=7.0cm]{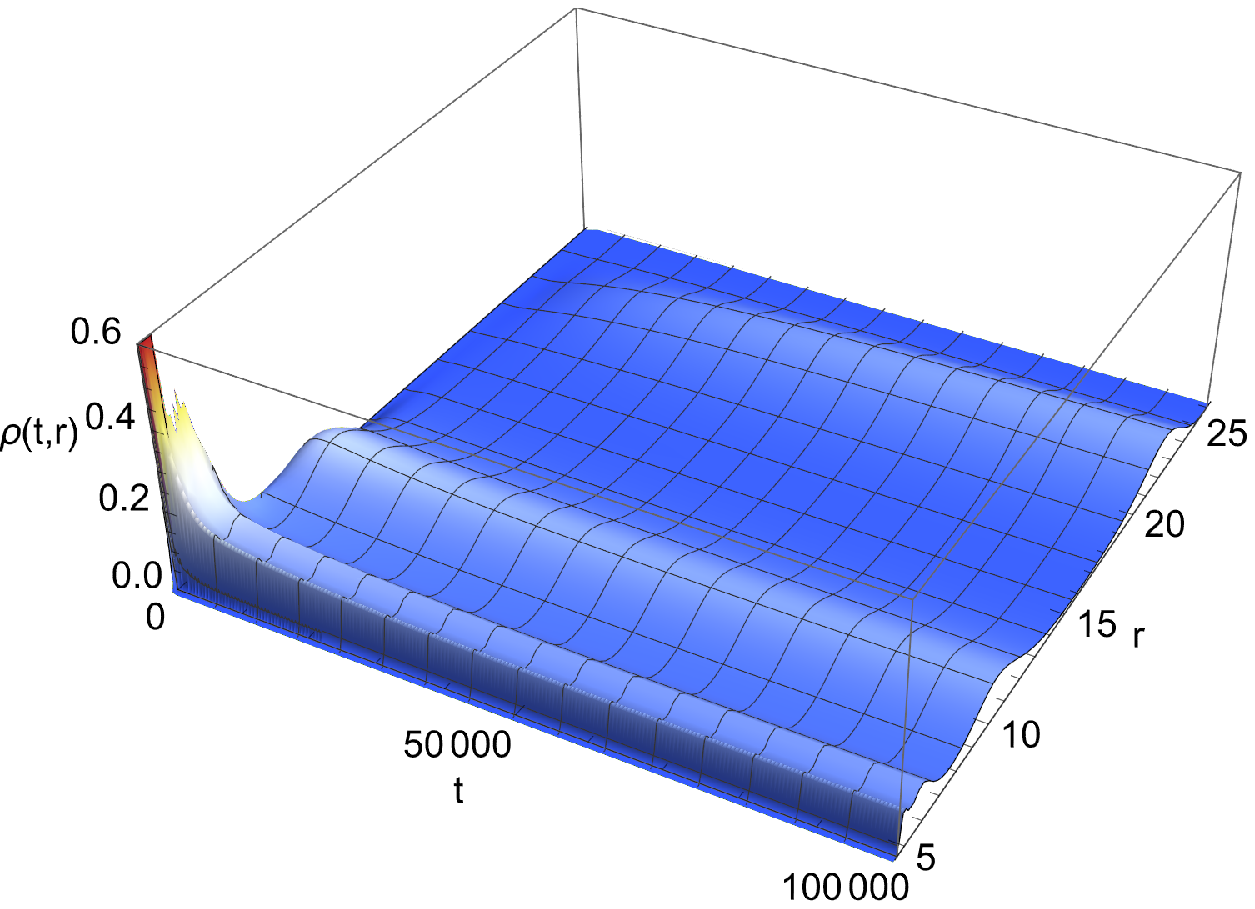}}}
\caption{Time evolution of the probability distribution $\rho(t, r)$ at middle ensemble temperature $T_{\rm E}=0.00631913$. The initial Gaussian wave packet is set to be peaked at the coexistent (a) largest (b) large (c) small and (d) smallest black hole states.}\label{pprhotc3}
\end{figure}
\end{center}

\section{Dynamic processes at the quadruple point}$\quad$\\

  In what follows, we shall denote the horizon radius  by $r$ for simplicity. The probability that the system will stay in a particular black hole state with horizon radius $r$ is represented by a distribution denoted by $\rho(t, r)$. Based on the free energy landscape, black hole systems can undergo thermodynamic phase transitions and change to other black hole states through a thermal fluctuation of the order parameter $r$. This  process can be described by  the Smoluchowski equation, which is a particular case of the Fokker-Planck equation \cite{Zwanzig},
\begin{eqnarray}
 \frac{\partial \rho(t, r)}{\partial t}=D
 \frac{\partial}{\partial r}\left(e^{-\beta G_{\rm L}(r)}
 \frac{\partial}{\partial r}\left(e^{{\beta G_{\rm L}(r)}}\rho(t, r)\right)\right) \label{FPE}
\end{eqnarray}
 where
$D=k_{\rm B}T_{\rm E}/\zeta$ is the diffusion coefficient and $k_{\rm B}$ and $\zeta$ are Boltzmann's constant and the dissipation coefficient, respectively.

We shall use the Smoluchowski equation  to study the dynamic evolution of different black hole phases at a quadruple point. Without loss of generality, we work in  units where  $k_{\rm B}=\zeta=1$. To solve this partial differential equation, we need to impose both initial conditions and boundary conditions. For initial conditions, we set the initial  state to be some Gaussian wave packet 
\begin{eqnarray}\label{rhoinit}
 \rho(0, r)=\frac{1}{\sigma\sqrt{\pi}}e^{-\frac{(r-r_{{i} })^2}{\sigma^2}}
\end{eqnarray}
peaked at the horizon radius   $r_{{i}}$ corresponding to one
of the four distinct stable black hole phases.

We numerically set $\sigma=0.2$ 
and consider the initial state to be respectively peaked at  the  coexistent black hole state with smallest, small, large, and largest horizon radius, i.e. $r_{{i} }$=$r_{ ss}$,$r_{ s}$, $r_{ l}$, or $r_{ ll}$. 
We impose reflective boundary conditions at $r=0$ and $r=\infty$, where there are extremely high potential barriers
\begin{eqnarray}
e^{-\beta G_{\rm L}(r)}
 \frac{\partial}{\partial r}\left(e^{{\beta G_{\rm L}(r)}}\rho(t, r)\right)\bigg|_{r=r_{bdy}}=0\label{FPE}
\end{eqnarray}
to ensure that the total  probability is preserved over time.

\begin{center}
\begin{figure}
\center{\subfigure[]{\label{tmill2a}
\includegraphics[width=7.0cm]{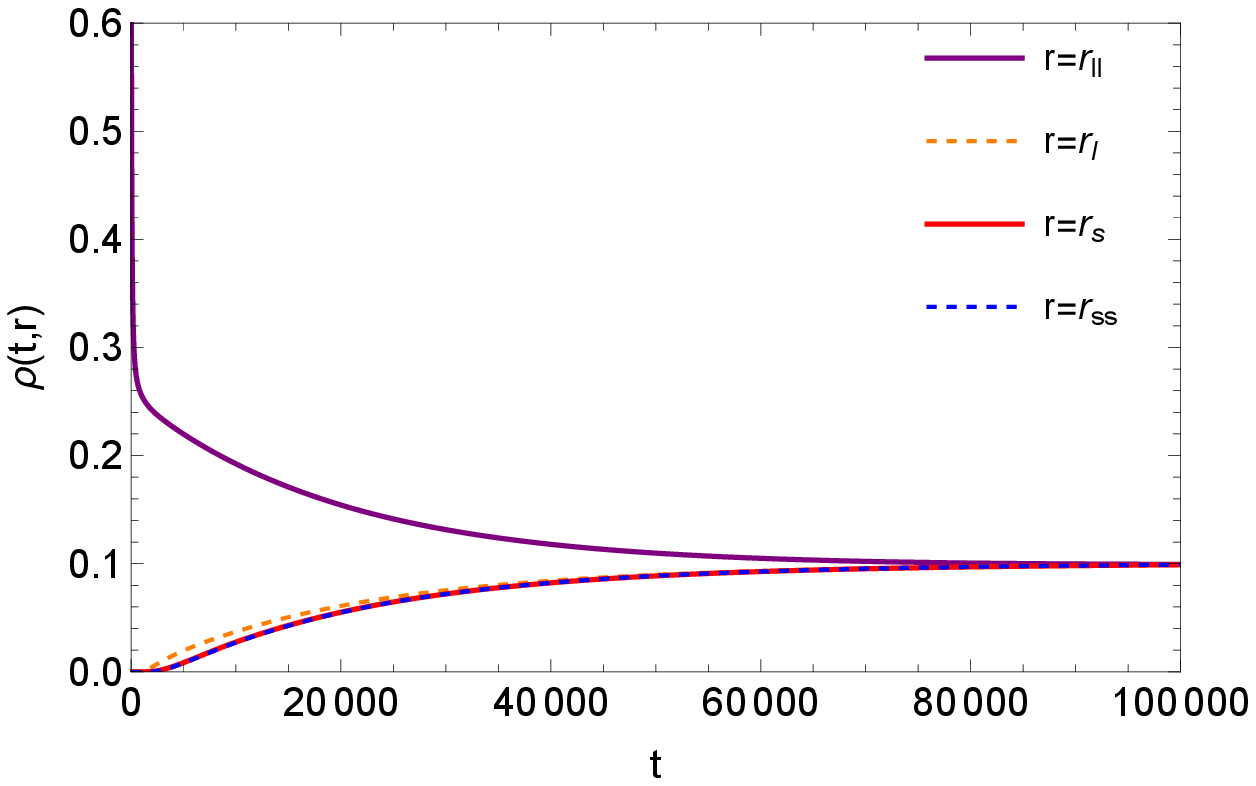}}
\subfigure[]{\label{tmil2b}
\includegraphics[width=7.0cm]{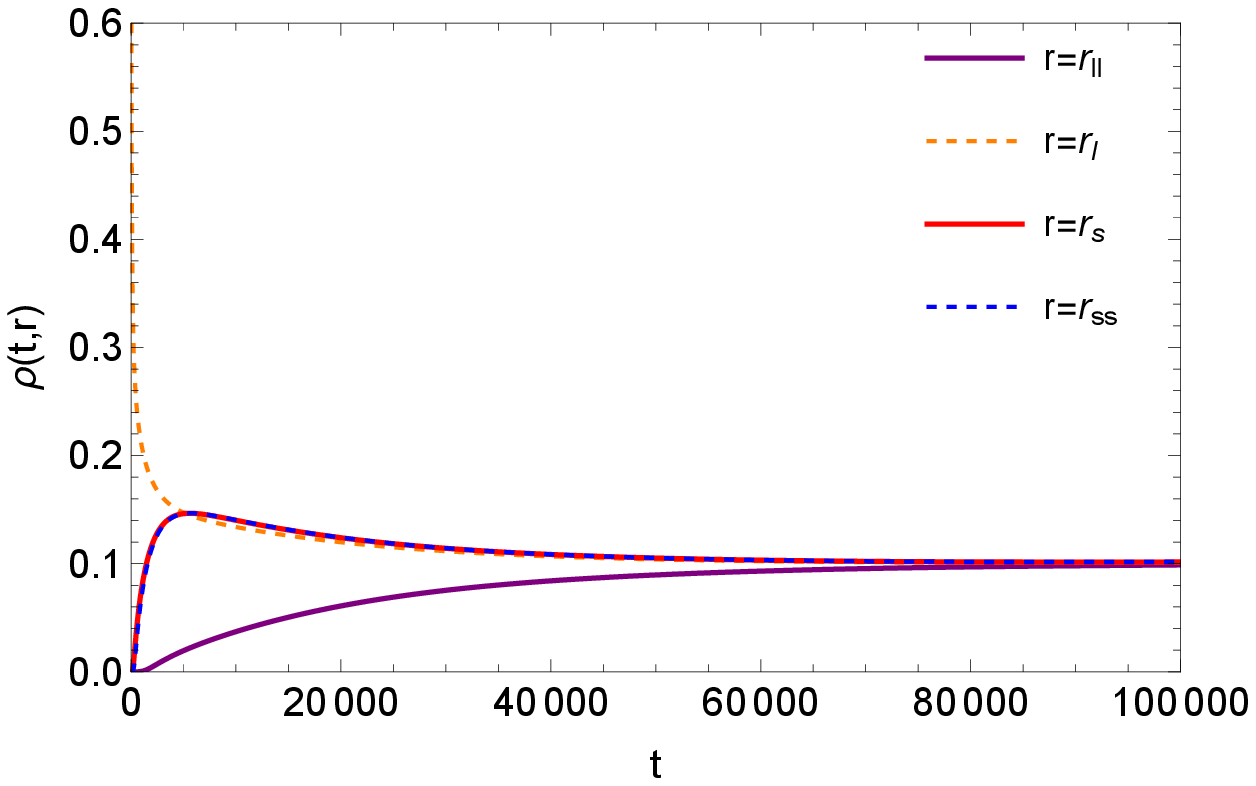}}\\
\subfigure[]{\label{tmis2c}
\includegraphics[width=7.0cm]{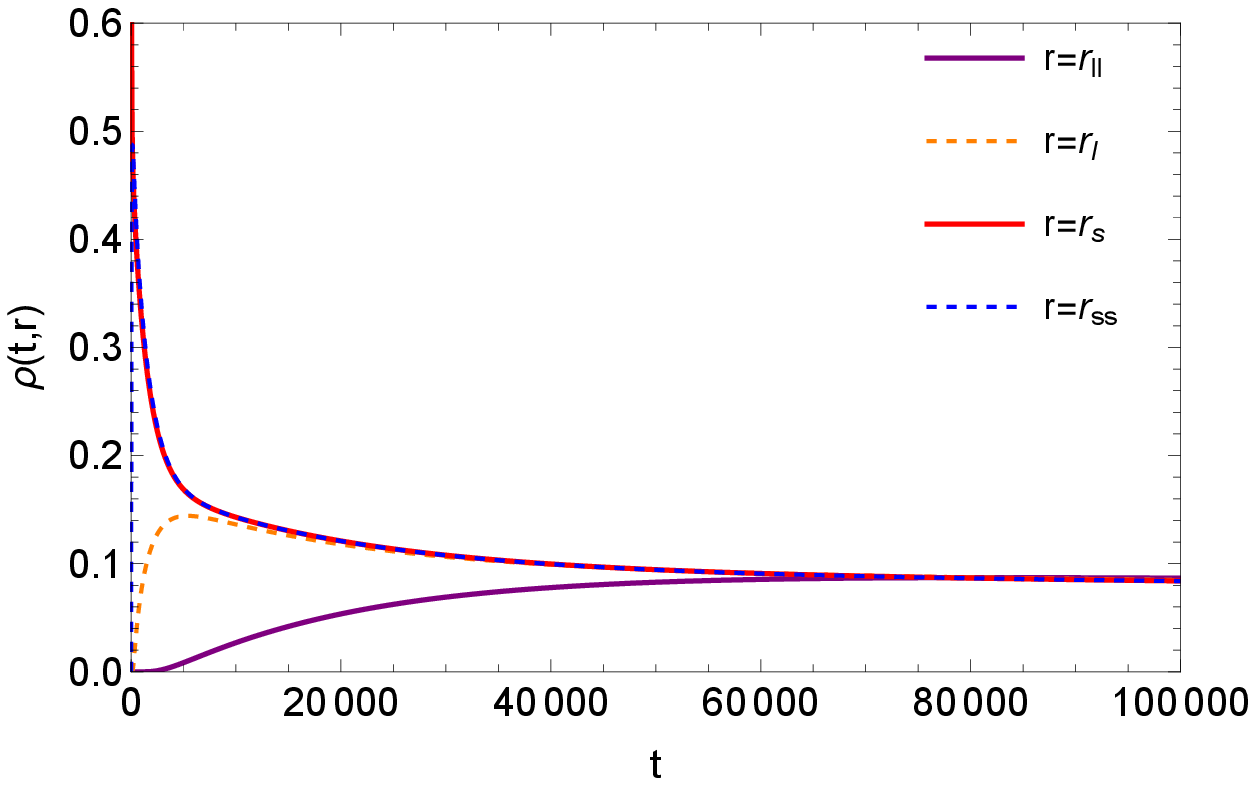}}
\subfigure[]{\label{tmiss2d}
\includegraphics[width=7.0cm]{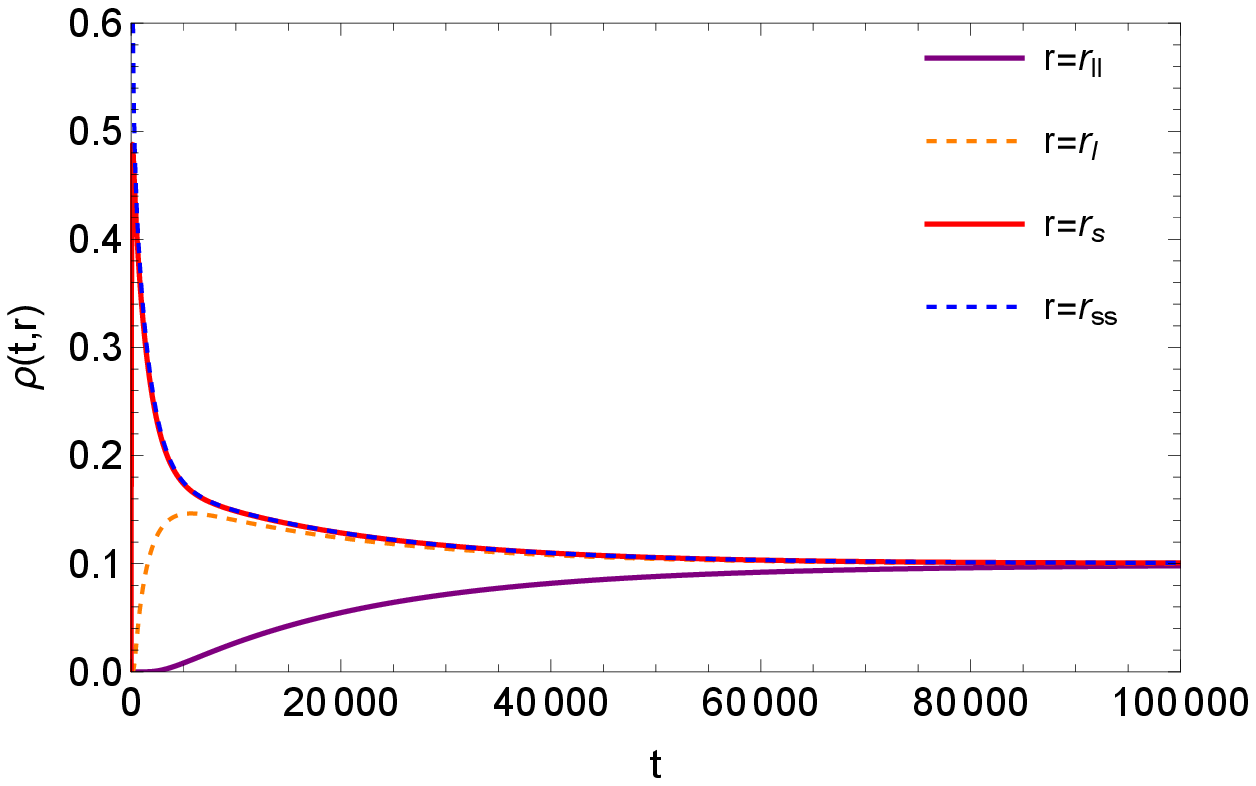}}}
\caption{Behaviours of $\rho(t, r)$ for four stable phases at middle ensemble temperature $T_{\rm E}=0.00631913$. The initial Gaussian wave packet is set to be peaked at the coexistent (a) largest (b) large (c) small and (d) smallest black hole states.}\label{pprhotc4}
\end{figure}
\end{center}

We numerically choose the left boundary and right boundary to be at $r=1$ and $r=30$ to solve the Smoluchowski equation.  The final stationary probability distributions, $\rho(t, r)$ should not depend on time  (the probability current must vanish),
 so the final state  probability distributions  $   \rho(r)\propto e^{-\beta G_{\rm L}(r)}$ must only depend on the structure of Gibbs free energy, which is determined by the ensemble temperature.

We consider the behaviour of $G_{\rm L}$ for three distinct values of the ensemble temperature $T_{\rm E}$, shown in Fig.~\ref{Gibbs}.
 At the intermediate ensemble temperature (Fig.~\ref{ga}) the four local minima representing four stable states in the Gibbs free energy  diagram share the same value $G_{\rm L}=5.50326$. For a slightly smaller value of $T_{\rm E}$ 
 (Fig.~\ref{gb}) the minima are no longer degenerate, and instead increase from 
$r_{ ss}$ to $r_{ ll}$.
For a slightly larger value of $T_{\rm E}$ 
 (Fig.~\ref{gc}) the minima decrease from 
$r_{ ss}$ to $r_{ ll}$.\\

\begin{center}
\begin{figure}[h]
\center{\subfigure[]{\label{rhota}
\includegraphics[width=7.0cm]{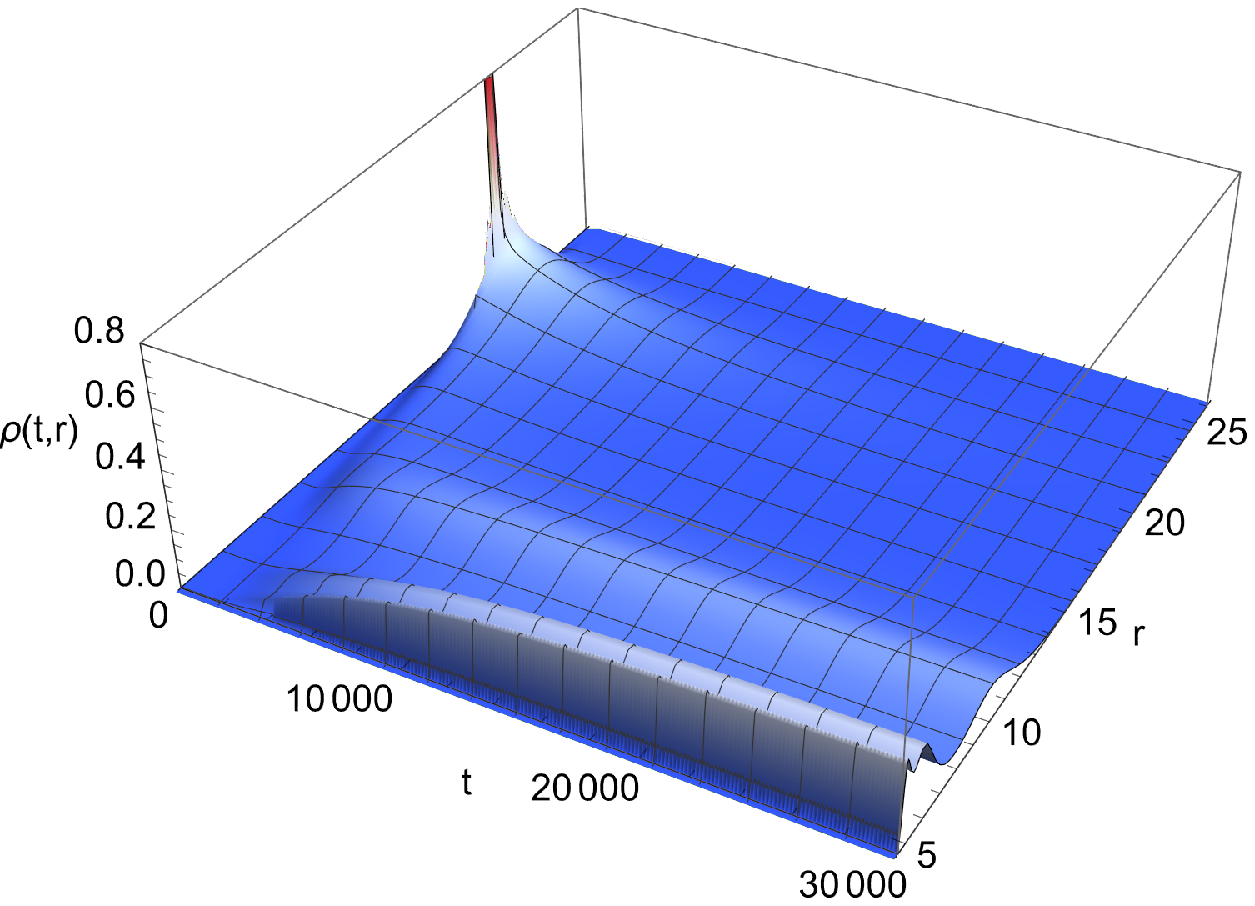}}
\subfigure[]{\label{rhotb}
\includegraphics[width=7.0cm]{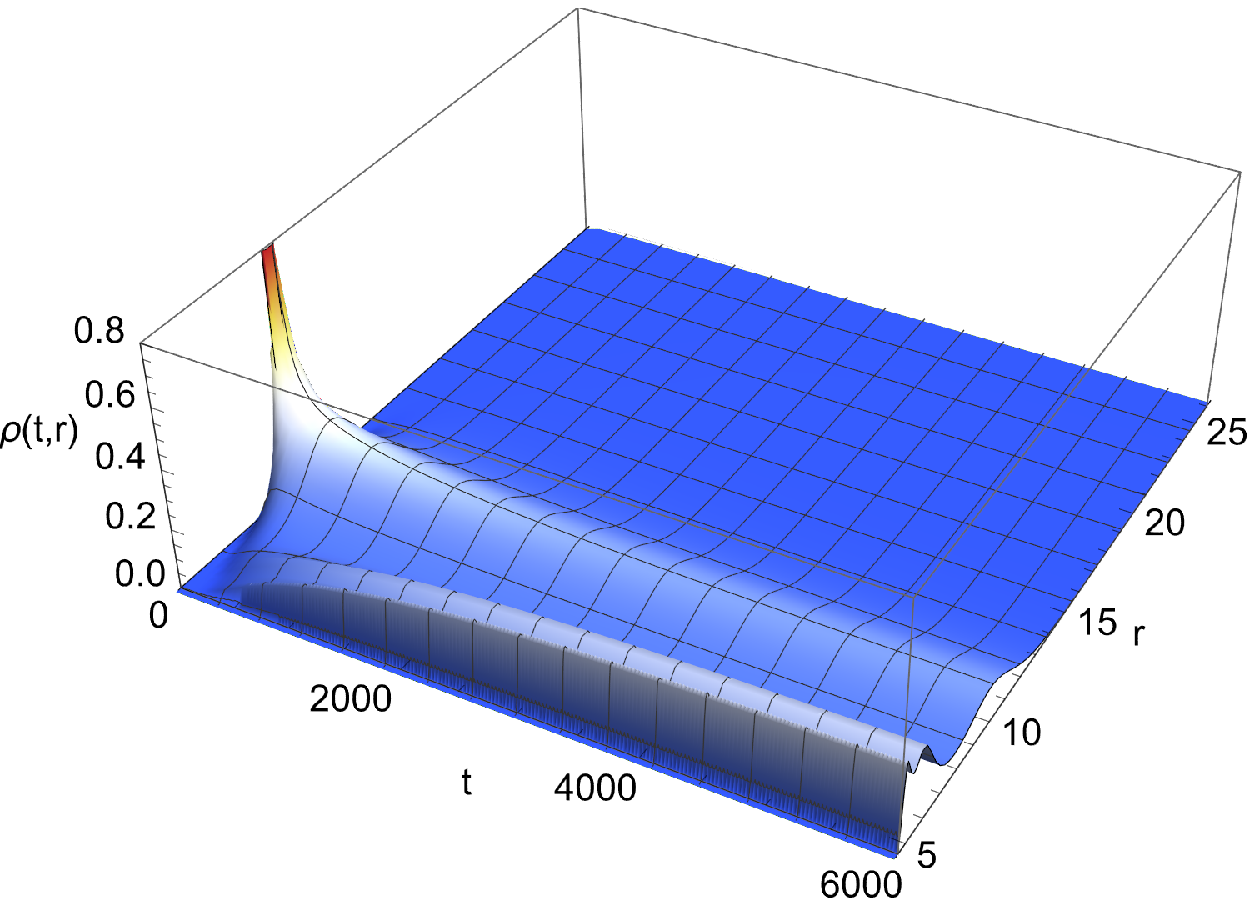}}\\
\subfigure[]{\label{rhotc}
\includegraphics[width=7.0cm]{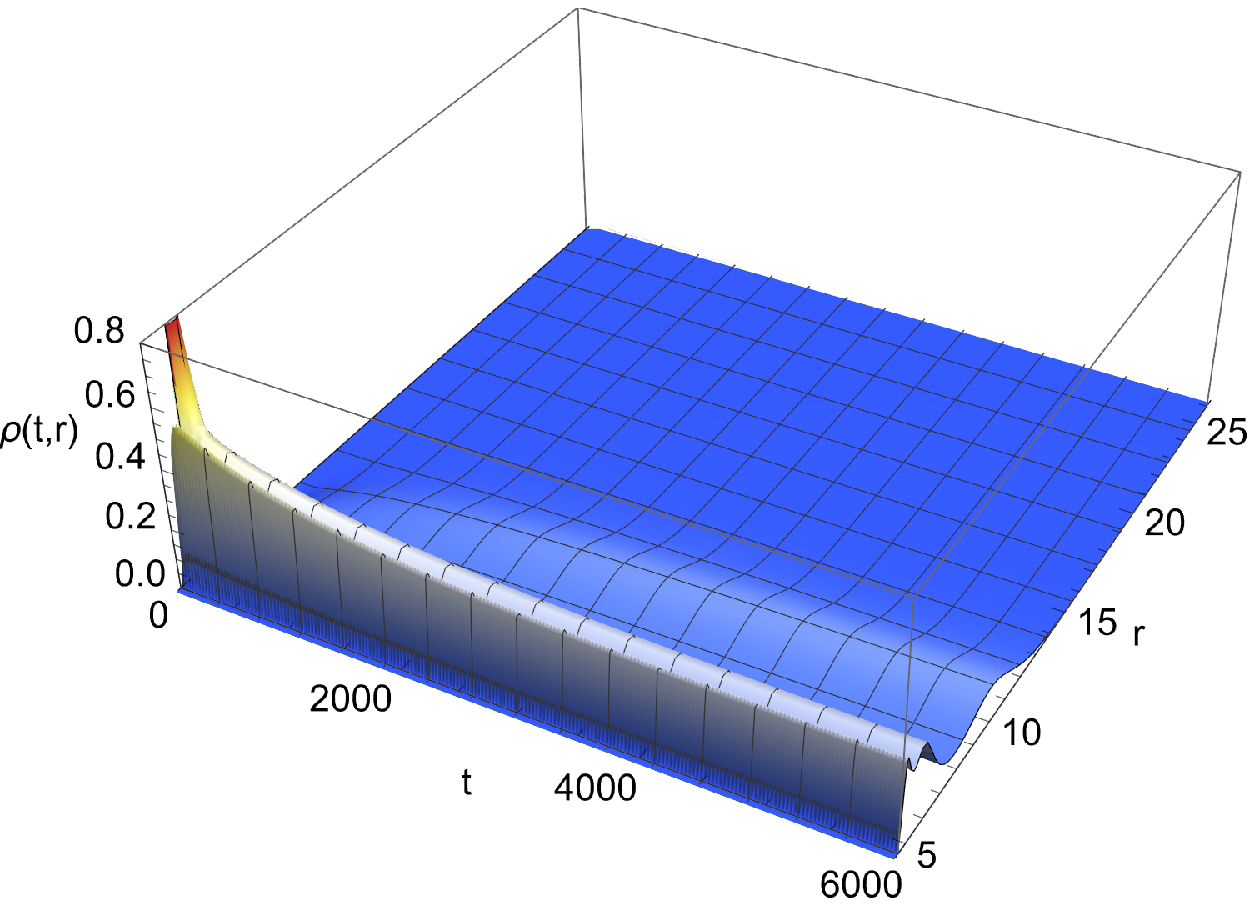}}
\subfigure[]{\label{rhotd}
\includegraphics[width=7.0cm]{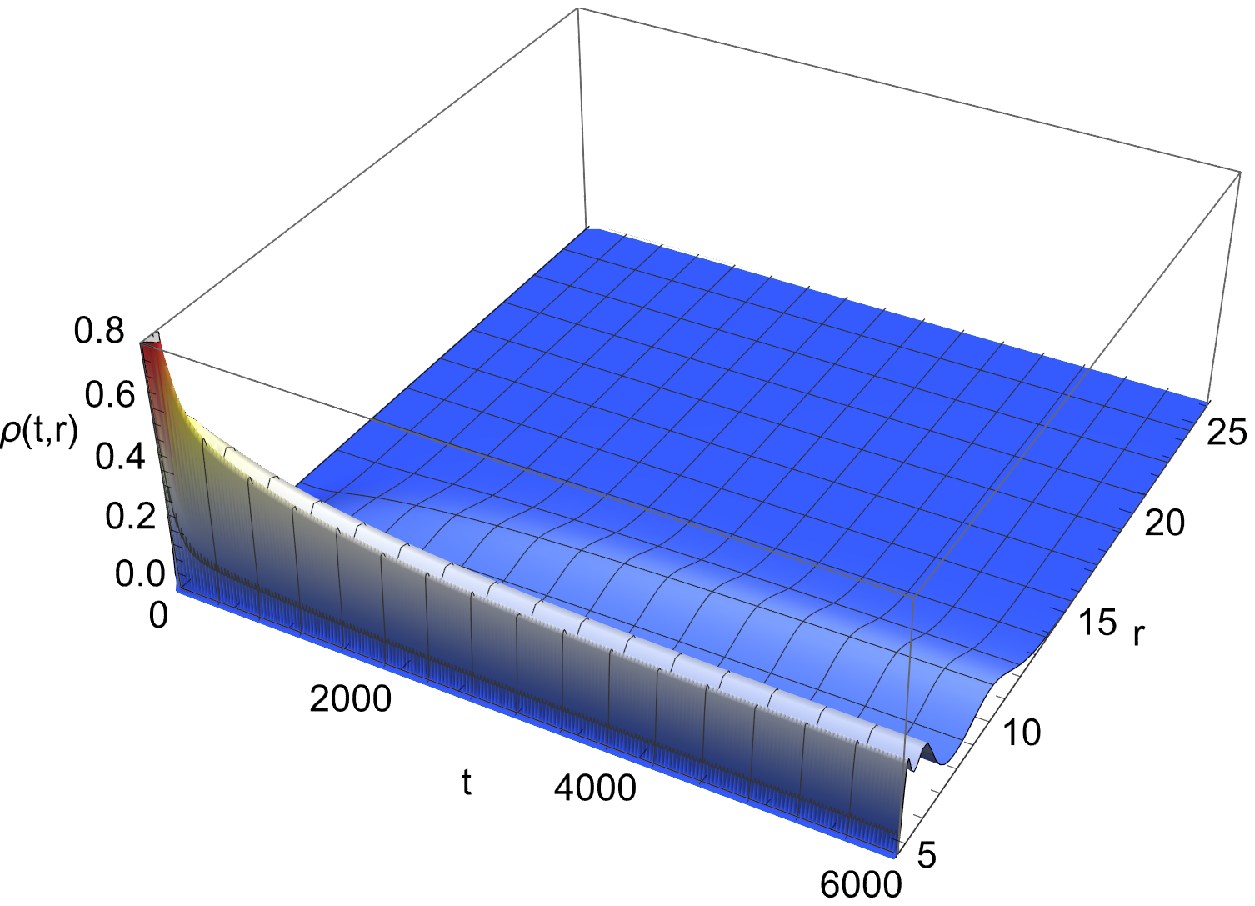}}}
\caption{Time evolution of the probability distribution $\rho(t, r)$ at low ensemble temperature $T_{\rm E}=0.006298$. The initial Gaussian wave packet is set to be peaked at the coexistent (a) largest (b) large (c) small and (d) smallest black hole states.}\label{lowrho}
\end{figure}
\end{center}

\end{widetext}

\subsection{Intermediate Ensemble Temperature}

 We plot  in Fig.~\ref{pprhotc3} the behaviour of the probability distribution $\rho$ as a function of $(t,r)$, for initial states peaked at each of the $r_{ i }$=$r_{ ll}$, $r_{ l}$, $r_{ s}$, and $r_{ ss}$. In  Fig.~\ref{pprhotc4} 
we plot for each of these successive cases the evolution of $\rho$ 
at each of the values $r_{i}$=$r_{ ss}$, $r_{ s}$, $r_{ l}$, and $r_{ ll}$.
 
 Consider the first case, shown in
Fig.~\ref{tmill} and
 Fig.~\ref{tmill2a}, in which   the probability is initially peaked at the black hole state with largest horizon radius, which means $\rho(0, r_{ ll})\gg\rho(0, r_{ l})\approx \rho(0, r_{ s}) \approx \rho(0, r_{ ss})\approx 0$. As time increases, there is a decline of the probability of the largest phase (with radius $r_{ ll}$), whilst the probabilities of the other three phases gradually increase. Note that in this process  $\rho(t, r_{ l})>\rho(t, r_{ s})> \rho(t, r_{ ss})$ with $\rho(t, r_{ s})\approx \rho(t, r_{ ss})$. This is because the off-Shell Gibbs free energy, interpreted as a potential, has more barriers between the largest and the two smallest black hole phases,   with radii $r_{ s}$ and $ r_{ ss}$ than the phase with radius $r_{ l}$. The reason that  $\rho(t, r_{ s})\approx \rho(t, r_{ ss})$ is that the barrier width (defined as the separation between  a minimum and the adjacent maximum)
 and height is relatively small between these two black hole states. After a sufficiently long time, say $t>100000$, the four states equilibrate to the multicritical state, sharing the same probability $\rho(r_{ {k}}, t)\approx$0.1, for ${k} =$ $ll, l, s$ and $ss$, due to the degenerate local minima value  in the Gibbs free energy  diagram
 Fig.~\ref{ga}.

Consider the next case shown in Fig.~\ref{tmil} and \ref{tmil2b}, in which the initial wave packet is set to be located at the large black hole state with radius $r_{ l}$. At the beginning, we have  $\rho(0, r_{ l})\gg\rho(0, r_{ ll})\approx \rho(0, r_{ s}) \approx \rho(0, r_{ ss})\approx 0$.  At early times, the probability of the phase with large radius declines rapidly, whilst $\rho(t, r_{ ll}), \rho(t, r_{ s}) $, and $ \rho(t, r_{ ss})$ increase. It is  noteworthy that  the probability leakage  to the  small and smallest black hole states
grows much faster than 
that to the largest state. This is 
 due to the relatively small potential barrier heights and widths
between the large state and the two smallest states. 
 The probabilities
$\rho(t, r_{ s})$, and $\rho(t, r_{ ss})$ quickly reach their maximal values, which are even 
 (slightly) 
larger than the probability  $\rho(t, r_{ l})$ of the given initial state.  This 
signifies a change of dominant state, in this case
 from the large phase  to the small and smallest black hole phases. This  phenomenon is called strong oscillatory behaviour \cite{Wei:2021bwy}. At the end, the four probabilities become stationarity and equilibrate to almost   the same value,  as expected due to the structure of $G_{\rm L}$.

\begin{widetext}
\begin{center}
\begin{figure}[h]
\center{\subfigure[]{\label{tlowll2a}
\includegraphics[width=7.0cm]{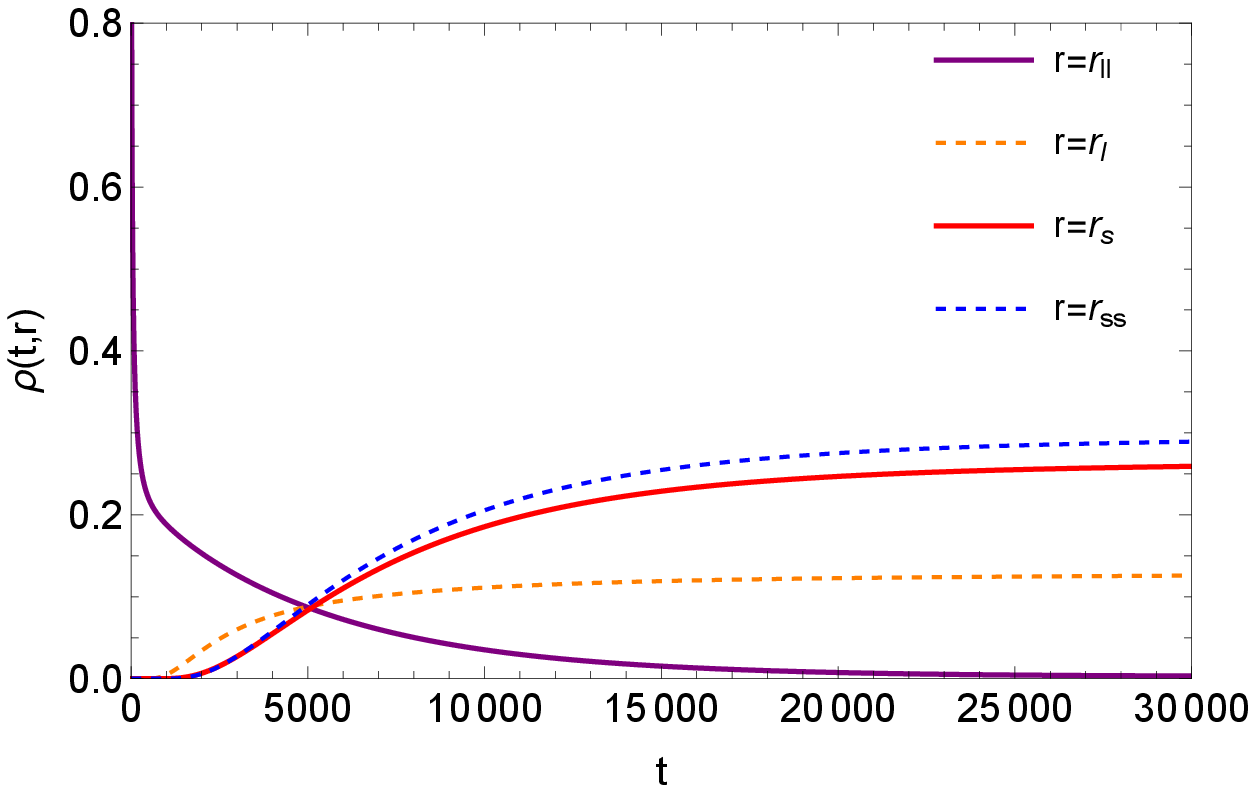}}
\subfigure[]{\label{tlowl2b}
\includegraphics[width=7.0cm]{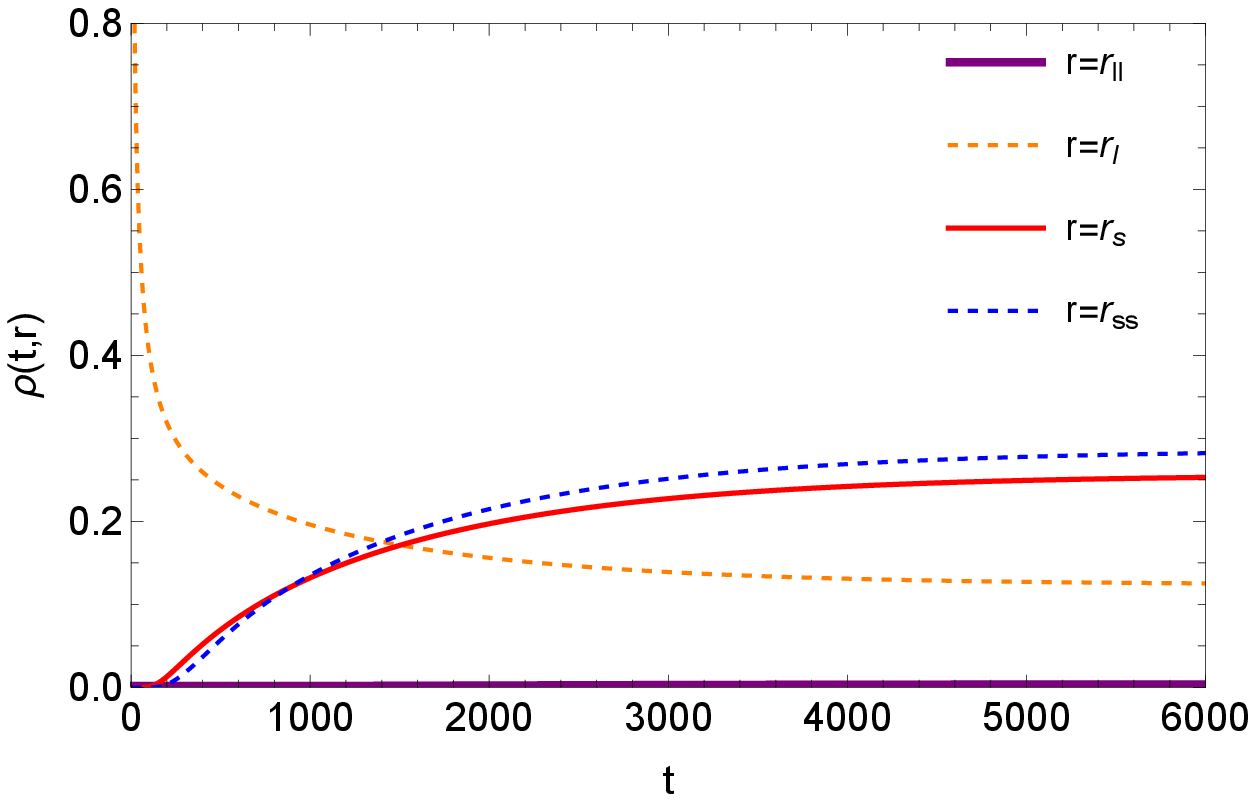}}\\
\subfigure[]{\label{tlows2c}
\includegraphics[width=7.0cm]{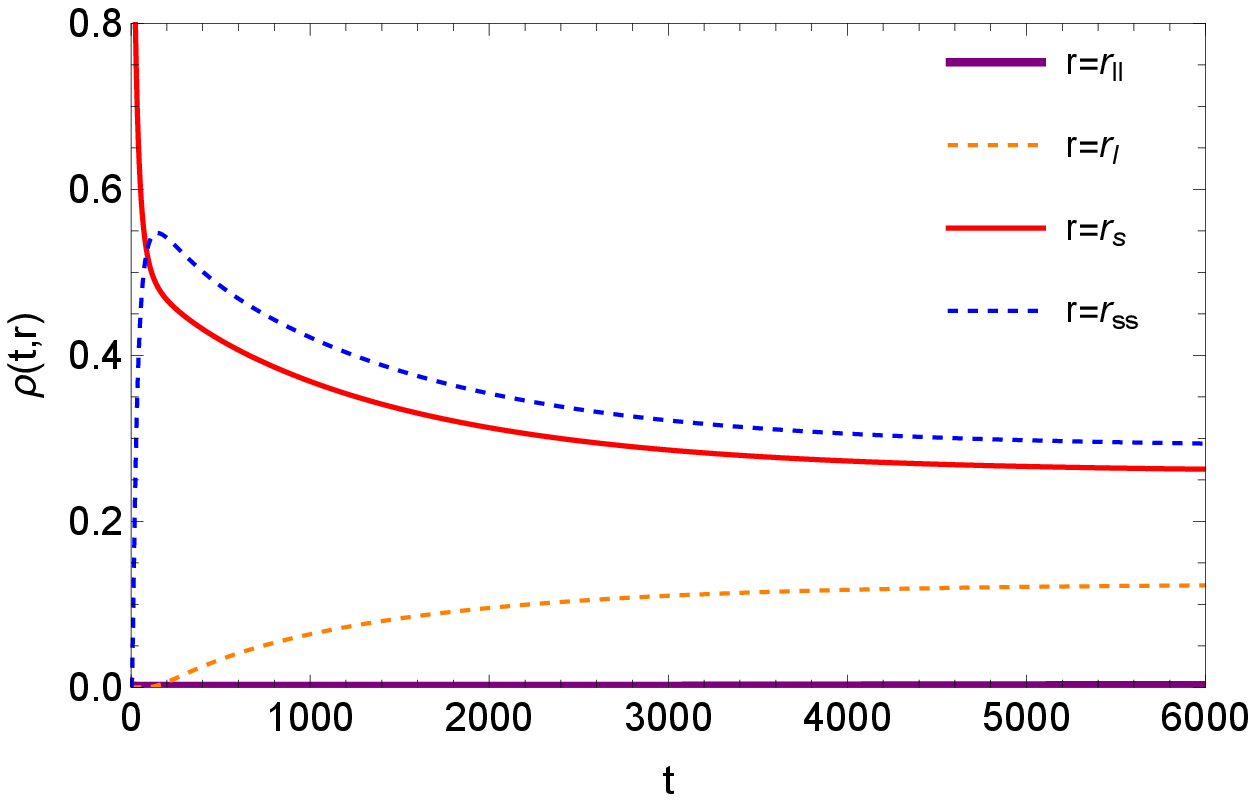}}
\subfigure[]{\label{tlowss2d}
\includegraphics[width=7.0cm]{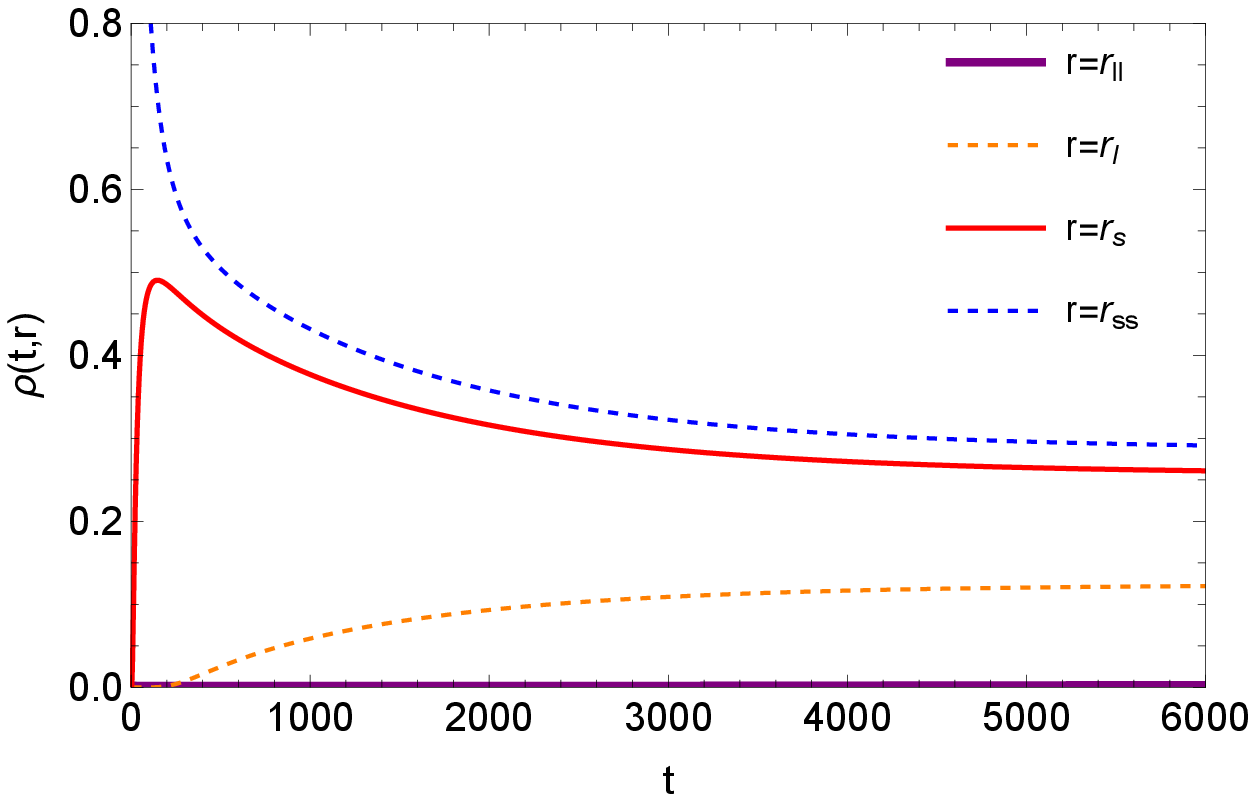}}}
\caption{Behaviours of $\rho(t, r)$  for four stable phases at low ensemble temperature $T_{\rm E}=0.006298$. The initial Gaussian wave packet is initialized to  peak at the coexistent (a) largest (b) large (c) small and (d) smallest black hole states.}\label{lowrho2d}
\end{figure}
\end{center}
 \end{widetext}

Turning now to the remaining two cases, consider the initial wave packet  peaked at $r=r_{s}$, as shown in  Fig.~\ref{tmis} and 
Fig.~\ref{tmis2c}. At early times, the probability leaks from $\rho(t, r_{ s})$ to $\rho(t, r_{ ss})$ very quickly since the potential barrier width and height between the small and smallest black hole states are extremely small. After a short   time,  the probability curves for the small and smallest black hole phases almost overlap. The probability  $\rho(t, r_{ l})$ for the large phase grows to a maximum value smaller than that of $\rho(t, r_{ s})$ and $\rho(t, r_{ ss})$, then decreases to the stationary value. This pheonomenon is called weak  oscillatory behaviour \cite{Wei:2021bwy}. 
 In the evolution process,  $\rho(t, r_{ ll})$ for the largest black hole state has the lowest value and grows the slowest. This is because the largest phase is the furthest away from the initial state.

Finally, if the distribution is peaked at the smallest phase (Fig.~\ref{tmiss2d}), the behaviour of all  four curves resembles those of Fig.~\ref{tmis2c}, the only distinction being that the roles $\rho(t, r_{ s})$ and $\rho(t, r_{ ss})$ play are interchanged.

\subsection{Low and High Ensemble Temperatures}

 For low ensemble temperature $T_{\rm E} = 0.006298 $, shown in Fig.~\ref{gb}, the key feature of the off-shell Gibbs free energy is $G_{\rm L}( r_{ ss})<G_{\rm L}( r_{ s})<G_{\rm L}( r_{ l})<G_{\rm L}( r_{ ll})$. Hence the smallest black hole phase will be the final dominant state regardless of the given initial state, with the final stationary state probability distributions being $\rho(t, r_{ ss})>\rho(t, r_{ s})>\rho(t, r_{ l})>\rho(t, r_{ ll})$, as shown in Fig.~\ref{lowrho} and Fig.~\ref{lowrho2d}. So we observe strong oscillatory phenomena when the initial state is peaked at $r_{ ll}$, $r_{ l}$, and $r_{ s}$, as respectively shown in Fig.~\ref{tlowll2a},  Fig.~\ref{tlowl2b}, and  Fig.~\ref{tlows2c}, while  Fig.~\ref{tlowss2d} exhibits weak oscillatory behaviour. The final probability for the largest black hole state is extremely small compared to the other three states due to its largest Gibbs free energy value. The final  probability distributions $\rho(t, r_{ ss}) $ and $\rho(t, r_{ s})$ are relatively approximate due to the smallest barrier size between these two states. More structures can be viewed with more discernment. For instance, as depicted in Fig.~\ref{tlowll2a}, the large black hole phase  probability grows faster than that of the small and smallest black hole phases at early times, since the large black hole phase is closer to the initial state peaked at $r_{ ll}$. But the lower free energy of phases at $r=r_{ s}$  and $r=r_{ ss}$ reverses the situation after $t\approx 5000$.

 \begin{widetext}
\begin{center}
\begin{figure}[h]
\center{\subfigure[]{\label{rhothia}
\includegraphics[width=7.0cm]{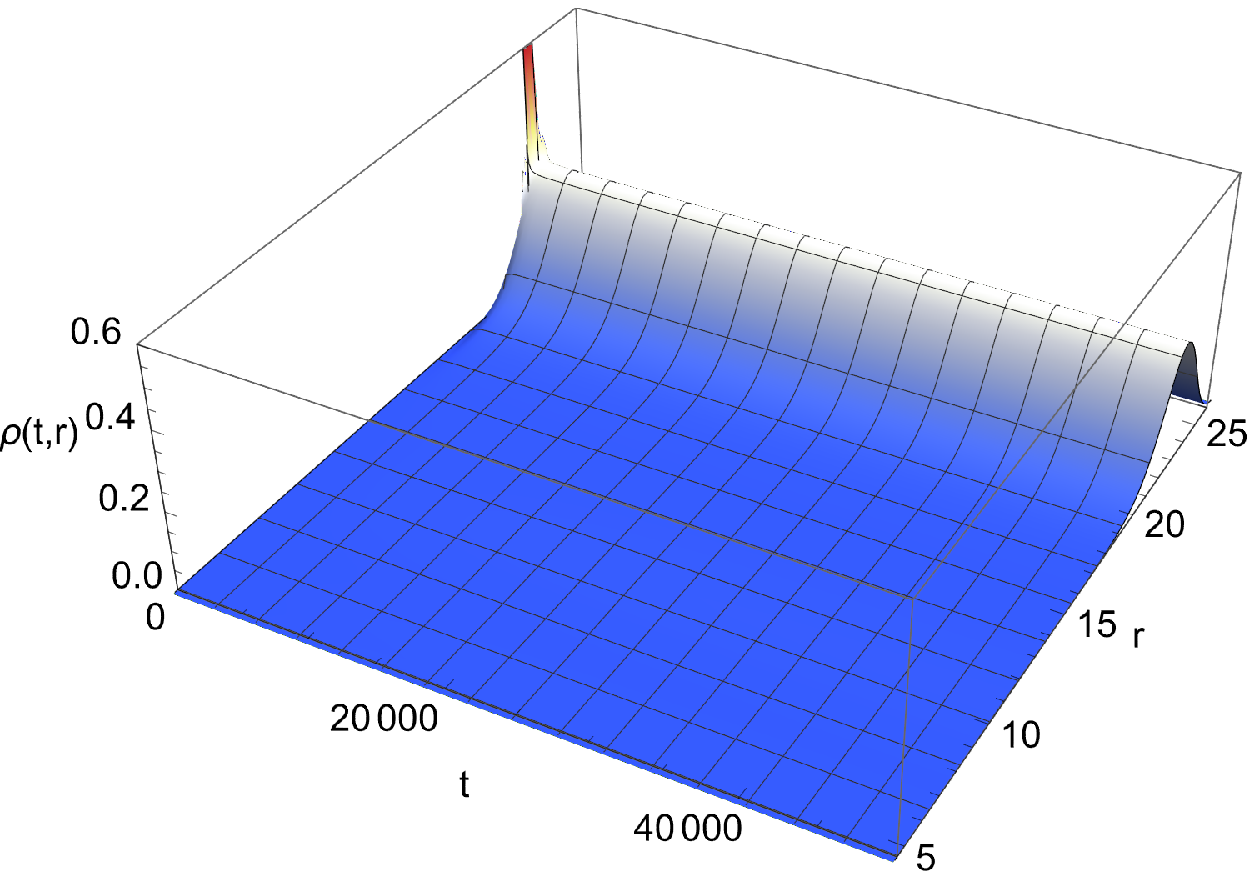}}
\subfigure[]{\label{rhothib}
\includegraphics[width=7.0cm]{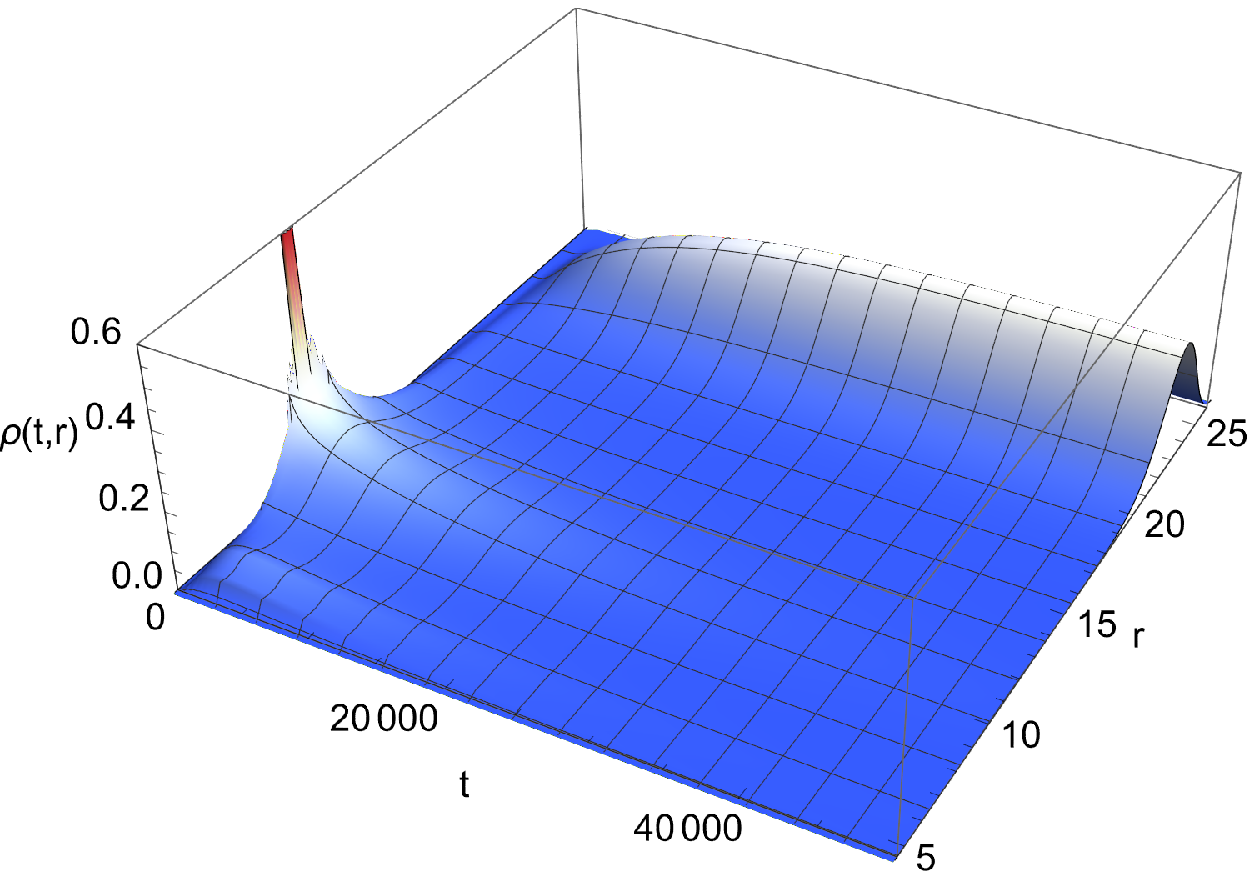}}\\
\subfigure[]{\label{rhothic}
\includegraphics[width=7.0cm]{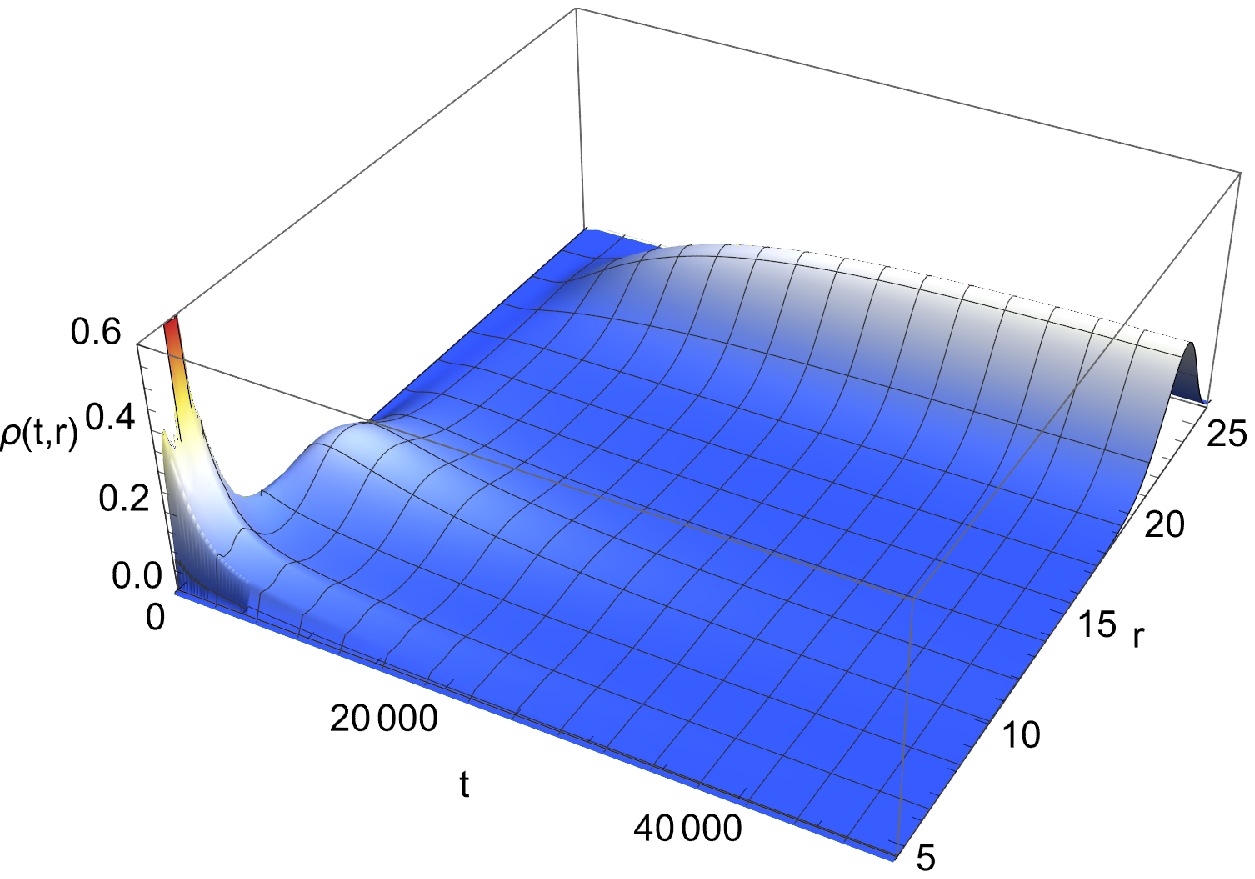}}
\subfigure[]{\label{rhothid}
\includegraphics[width=7.0cm]{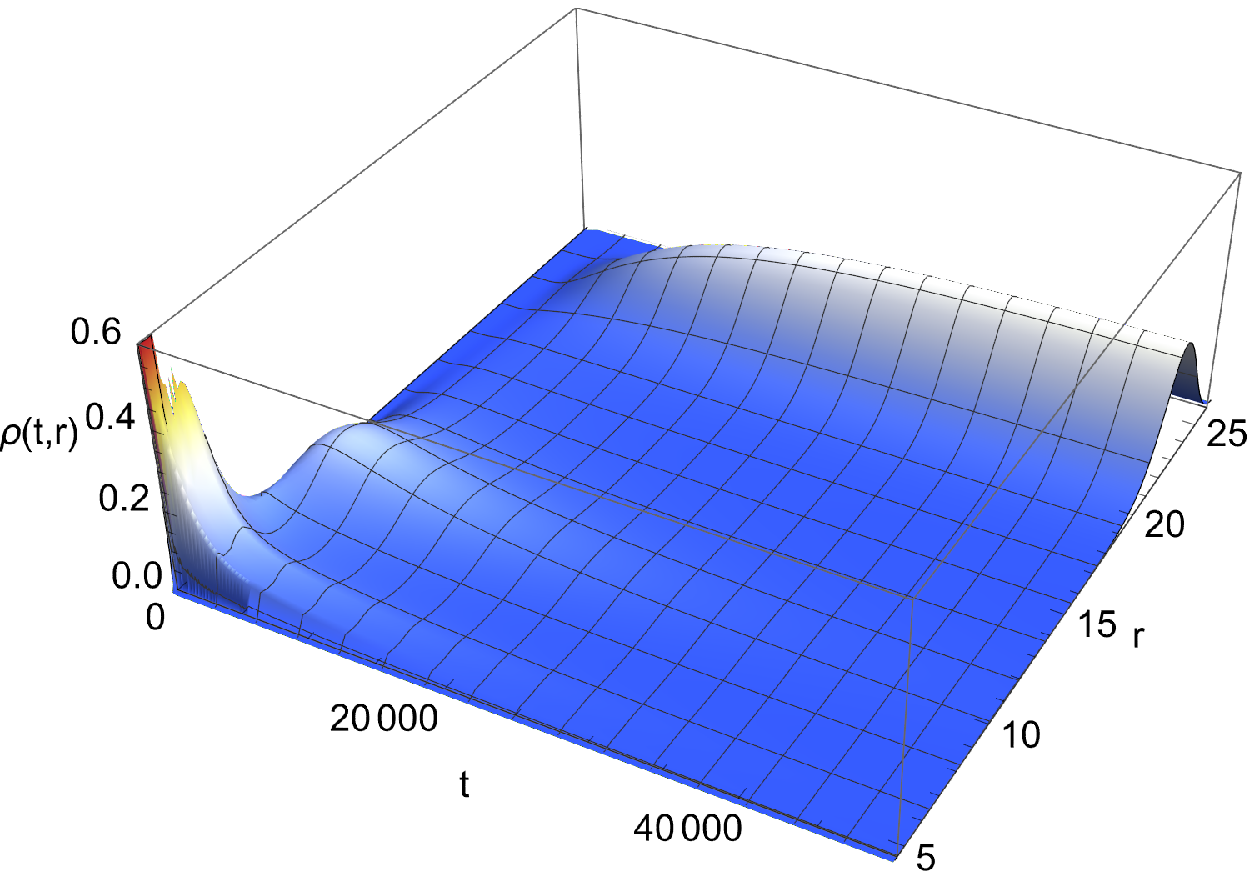}}}
\caption{Time evolution of the probability distribution $\rho(t, r)$ at high ensemble temperature $T_{\rm E}=0.0063422$. The initial Gaussian wave packet is initialized to peak at the coexistent (a) largest (b) large (c) small and (d) smallest black hole states.}\label{pprhotc}
\end{figure}
\end{center}

\begin{center}
\begin{figure}[h]
\center{\subfigure[]{\label{rhohi2a}
\includegraphics[width=7.0cm]{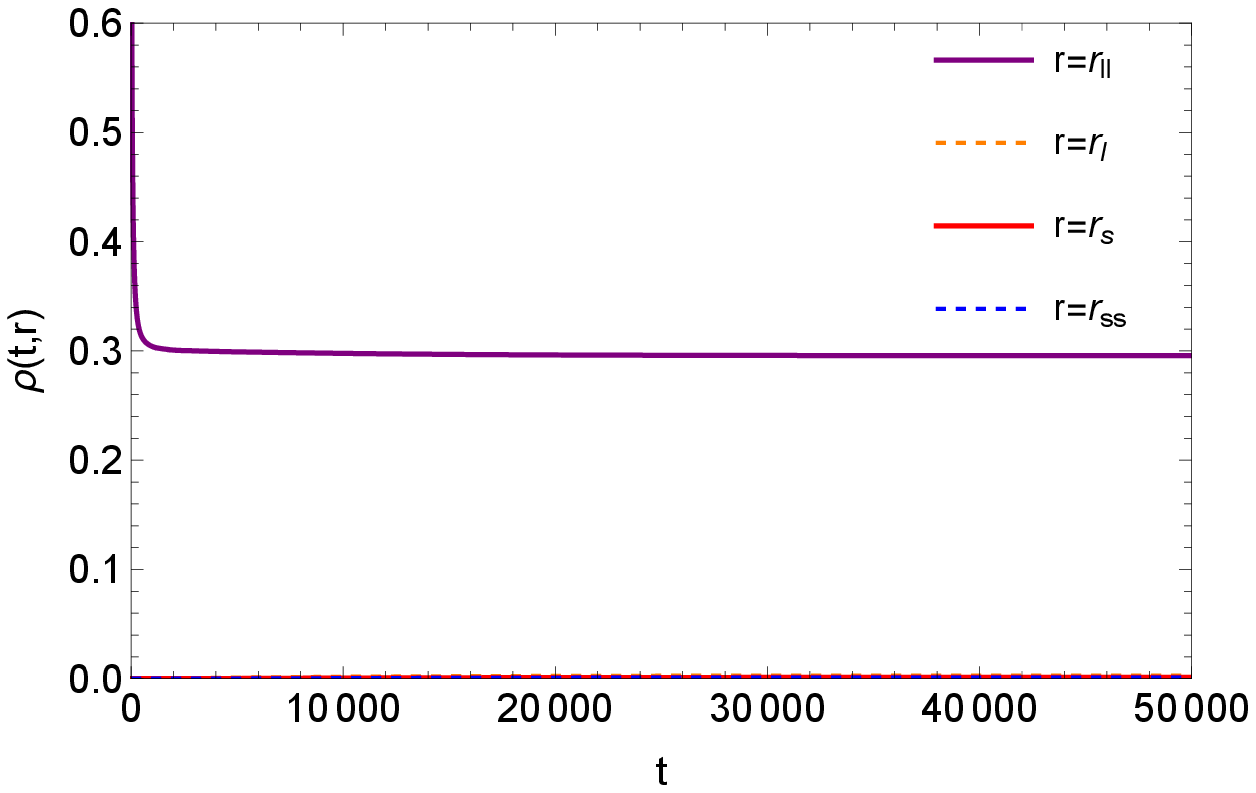}}
\subfigure[]{\label{rhohi2b}
\includegraphics[width=7.0cm]{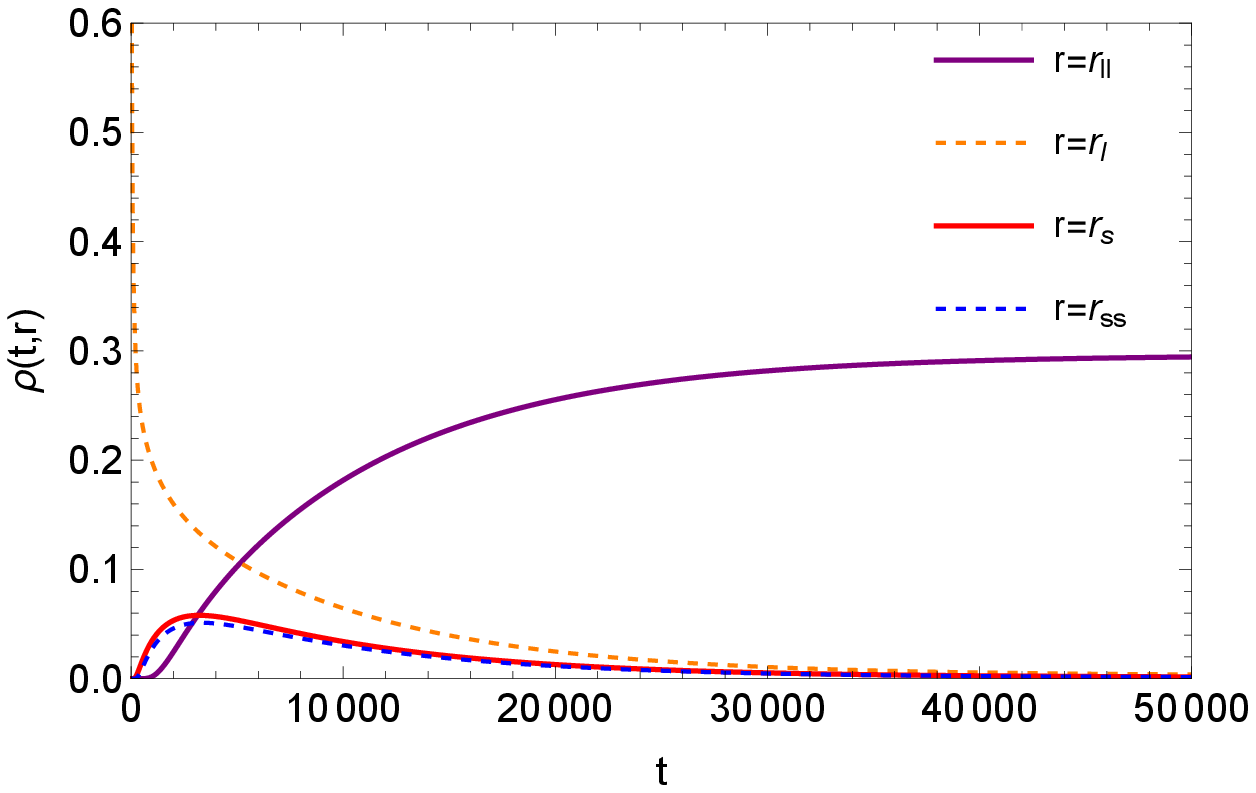}}\\
\subfigure[]{\label{rhohi2c}
\includegraphics[width=7.0cm]{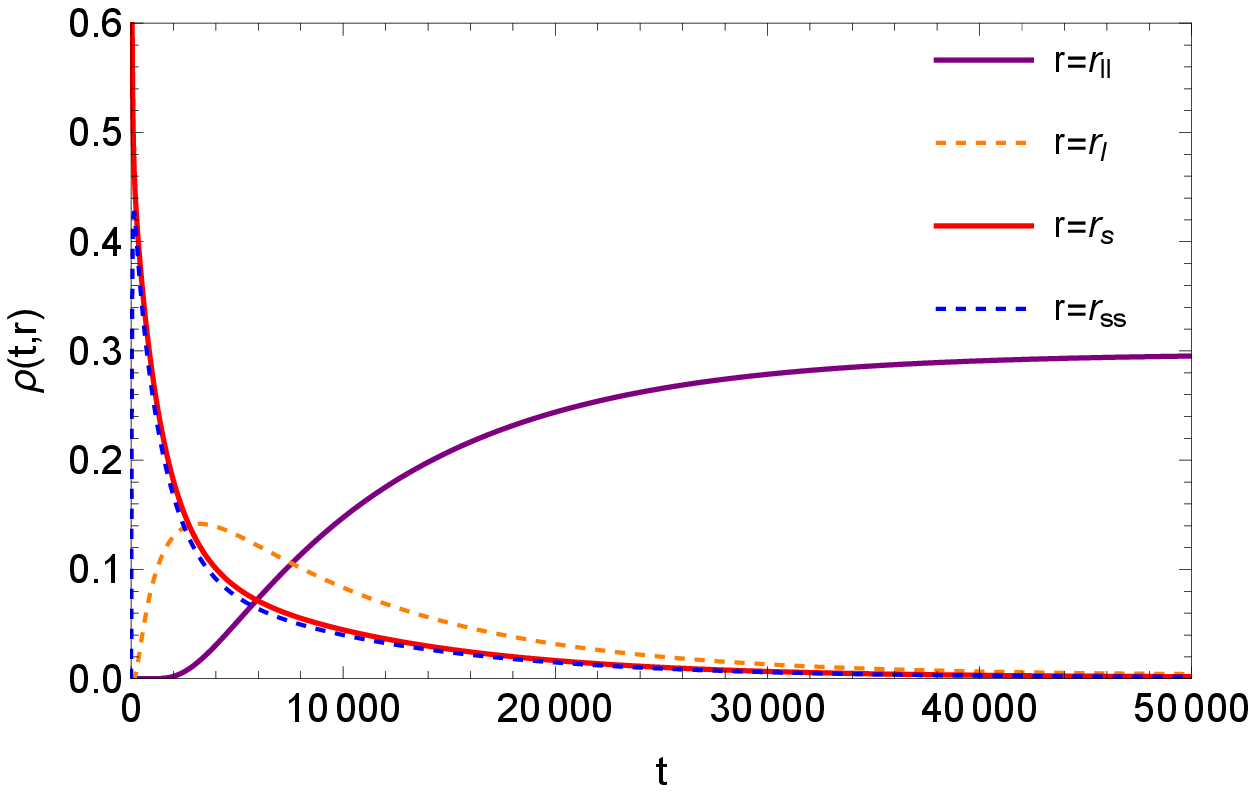}}
\subfigure[]{\label{rhohi2d}
\includegraphics[width=7.0cm]{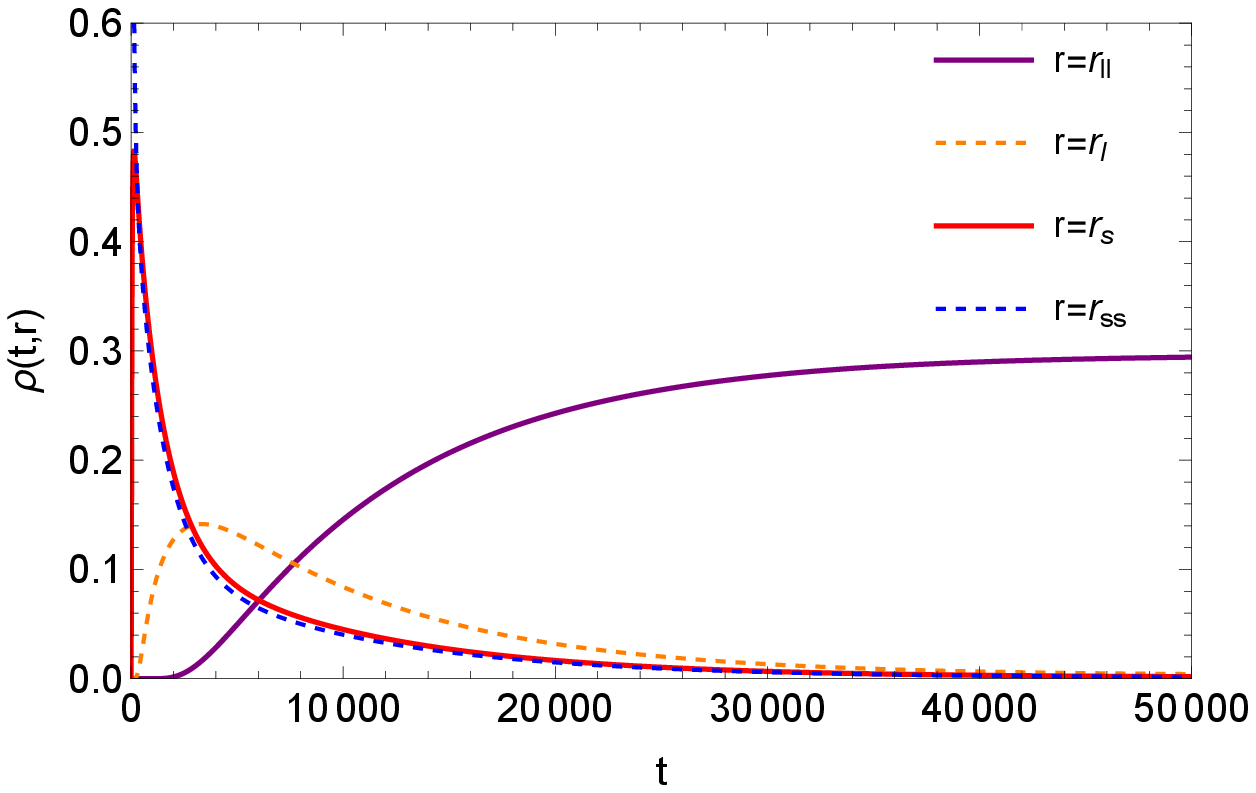}}}
\caption{Behaviours of $\rho(t, r)$  for four stable phases at the high ensemble temperature $T_{\rm E}=0.0063422$. The initial Gaussian wave packet is set to be peaked at the coexistent (a) largest (b) large (c) small and (d) smallest black hole states.}\label{high2d}
\end{figure}
\end{center}

 For high ensemble temperature $T_{\rm E} = 0.0063422 $, shown in Fig.~\ref{gc},    note that     $G_{\rm L}( r_{ ss})>G_{\rm L}( r_{ s})>G_{\rm L}( r_{ l})>G_{\rm L}( r_{ ll})$ in the $G_{\rm L}-r_+$ diagram,  and the barrier height and width to transit from the largest black hole state with radius $r_{ ll}$ to the other three states is extremely large. Hence, regardless of the given initial state, the largest black hole phase is  eventually dominant, as shown in Fig.~\ref{pprhotc}. Its  final probability is much larger than the other three phases,  with the final stationary states probability distribution $\rho(t, r_{ ll})\gg\rho(t, r_{ l})>\rho(t, r_{ s})>\rho(t, r_{ ss})$, illustrated in Fig.~\ref{high2d}. That is why strong oscillatory phenomena are observed in  Fig.~\ref{rhohi2b},  Fig.~\ref{rhohi2c}, and  Fig.~\ref{rhohi2d}. Taking a closer look at Fig.~\ref{rhohi2b}, initially $\rho(t, r_{ l})$ leaks to other three states and $\rho(t, r_{ s})\gtrsim \rho(t, r_{ ss})>\rho(t, r_{ ll})$  because the barrier widths amongst $ r_{ s},  r_{ ss}$ and $ r_{ l}$ are much smaller than that between $r_{ ll}$ and $ r_{ l}$ . Afterward $\rho(t, r_{ ss})\approx \rho(t, r_{ s})<\rho(t, r_{ ll})$ due to the lowest Gibbs value at $r=r_{ ll}$. Similar behaviours can be observed in  Fig.~\ref{rhohi2c}, and  Fig.~\ref{rhohi2d}. Hence, in the early stage, barrier widths and heights play an important role in the evolution process, whereas in the later stage, the value of the Gibbs free energy becomes crucial.

\section{ The first passage events at different ensemble temperatures}

In this section, we investigate the underlying reasons for black hole phase transition behaviours at quadruple points by inspecting the first passage events. On the free energy landscape, the first passage time for a given initial phase describes the required time to first ascend to the top of nearby the free energy barriers that correspond to unstable phases. Reflecting boundary conditions are still
imposed at $r = 0$ and $r=\infty$, and 
to depict the first passage event we impose absorbing boundary conditions at $r_{ m1}, r_{ m2}$, and $r_{ m3}$ for each unstable state:  
\begin{align}
    \rho(t,r_{m1})= \rho(t,r_{m2})= \rho(t,r_{m3})=0
\end{align}

Focusing on the Intermediate ensemble temperature shown in
Fig.\ref{tmill}, let us denote the  first passage time distributions by $F_{p1}(t)$, $F_{p2}(t)$, $F_{p3}(t)$ and $F_{p4}(t)$ for the 4 initial cases $r_{i}$=$r_{ ss}$, $r_{ s}$, $r_{ l}$, and $r_{ ll}$, respectively. Applying the Smoluchowski equation and boundary conditions, the first passage time distributions are expressed as
\begin{eqnarray}
 F_{p1}(t)&=&-\frac{\partial \rho(t, r_{ m1})}{\partial r},\\
 F_{p2}(t)&=&\frac{\partial \rho(t, r_{ m1})}{\partial r}-\frac{\partial \rho(t, r_{ m2})}{\partial r},\\
  F_{p3}(t)&=&\frac{\partial \rho(t, r_{ m2})}{\partial r}-\frac{\partial \rho(t, r_{ m3})}{\partial r},\\
 F_{p4}(t)&=&\frac{\partial \rho(t, r_{ m3})}{\partial r}.
\end{eqnarray}


We plot the first passage time distribution at three different ensemble temperatures, as shown in Fig.~\ref{FPI}, Fig.~\ref{FPL}, and Fig.~\ref{FPH}. For each first passage time distribution curve,  $F_p(t)$ reaches a peak after a short time and then decays rapidly. This indicates there is a large probability that the first passage event occurs in a very short time. This provides us with information for understanding the behaviour of $\rho(t, r)$ at early times.

In Table~\ref{ta1},  we list the height and width of each barrier at the three different ensemble temperatures.
At each  temperature, we find the extreme point $t$ of the first passage time distribution  for each case and list these values in Table~\ref{ta2}.
  The differences amongst the various barrier widths are larger than the differences amongst the barrier heights, so 
 barrier widths are more crucial in determining transit behaviour. We therefore use barrier
width $W$ to label the barrier size in the following context.
What we conclude from Table~\ref{ta1} and Table~\ref{ta2} is that the time corresponding to the peak of the first passage time distribution $t$ is highly correlated with barrier size.

Consider, for example, the  intermediate temperature $T_I$ and high temperature $T_H$. From $t_{ll3}$ to $t_{ss1}$, the first passage time peak $t$ decreases, i.e., $t_{ll3}>t_{l3}>t_{l2}>t_{s2}>t_{s1}>t_{ss1}$, and we have $W_{ll3}>W_{l3}>W_{l2}>W_{s2}>W_{s1}>W_{ss1}$  -- it is easier to surmount a barrier with small width. Likewise at  low temperature $T_L$, we have $t_{l3}> t_{ll3}>t_{l2}>t_{s2}>t_{s1}>t_{ss1}$, with corresponding barrier widths $W_{l3} > W_{ll3}>W_{l2}>W_{s2}>W_{s1}>W_{ss1}$. The interchange of the largest first passage time is accompanied by an interchange of the largest barrier width.


 To explain the dynamic behaviour of multicritical black hole phase transitions at early times, we make the assumption that the time needed from the barrier top to reach the adjacent local minimum is much shorter than the first passage time to climb to the barrier top from the initial local stable state \cite{LiWang}. Thus we obtain the first passage time scales from each initial black hole state to other stable states, symbolizing the difficulty of transitioning from one stable phase to another, as shown in Table~\ref{ta3}.  Let us focus on the intermediate ensemble temperature $T_{\rm E}=0.00631913$. First, for the initial black hole state with the largest horizon radius $r_{ll}$, we have $ t{(ll\to l)}< t{(ll\to s)}\lesssim t{(ll\to ss)}$ ($1593<1817\lesssim 1827$), which means that it is easier for the probability  to leak to the state with radius $r_{l}$. This is why in Fig.~\ref{tmill2a}, we observe the phenomenon that  
$\rho(t, r_{ l})>\rho(t, r_{ s})\gtrsim \rho(t, r_{ ss})$ at early times. For a large initial black hole state with radius $r_{l}$, we have the early time behaviour $\rho(t, r_{ s})\gtrsim \rho(t, r_{ ss})>\rho(t, r_{ ll})$ shown in Fig.~\ref{tmil2b}, 
which we can understand from the phase transition time scale relations   $t{(l\to s)}\lesssim t(l\to ss)<t(l\to ll)$. For a small initial black hole state, shown in Fig.~\ref{tmis2c}, we find that $t({s\to ss})<t(s\to l)<t(s\to ll)$, explaining why  $\rho(t, r_{ ss})$ is initially the fastest-growing probability and $\rho(t, r_{ ll})$  the slowest, with $\rho(t, r_{ ss})>\rho(t, r_{ l})>\rho(t, r_{ ll})$. Similarly, if the initial state is the smallest black hole, shown in Fig.~\ref{tmiss2d}, the reason why $\rho(t, r_{ s})>\rho(t, r_{ l})>\rho(t, r_{ ll})$ at early times is that $t(ss\to s)<t(ss\to l)<t({ss\to ll})$.

\begin{center}
\begin{figure}[!h]
\center{\subfigure[]{\label{IFP4a}
\includegraphics[width=7cm]{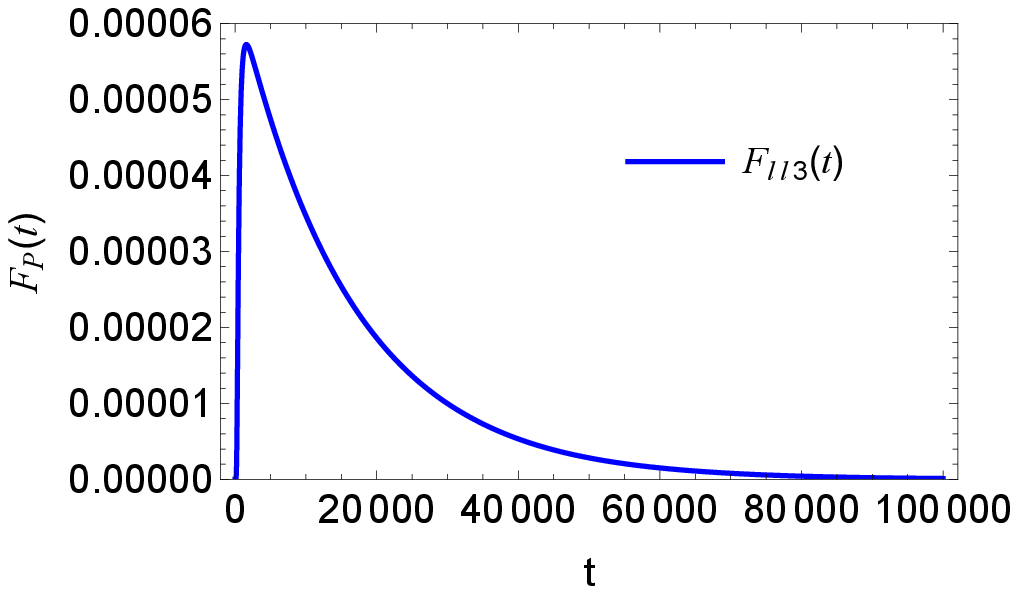}}
\subfigure[]{\label{FP4b}
\includegraphics[width=7cm]{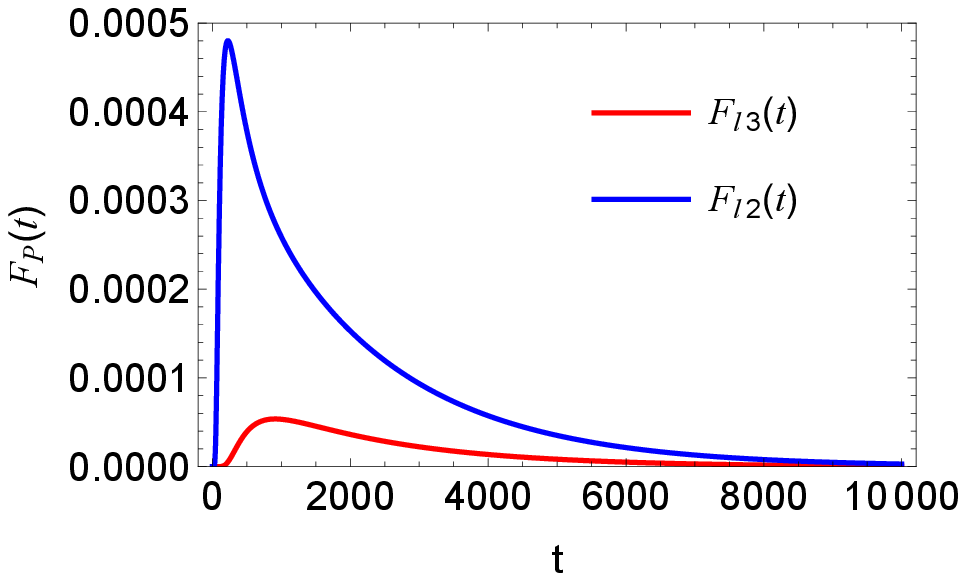}}}
\center{\subfigure[]{\label{FP4a}
\includegraphics[width=7cm]{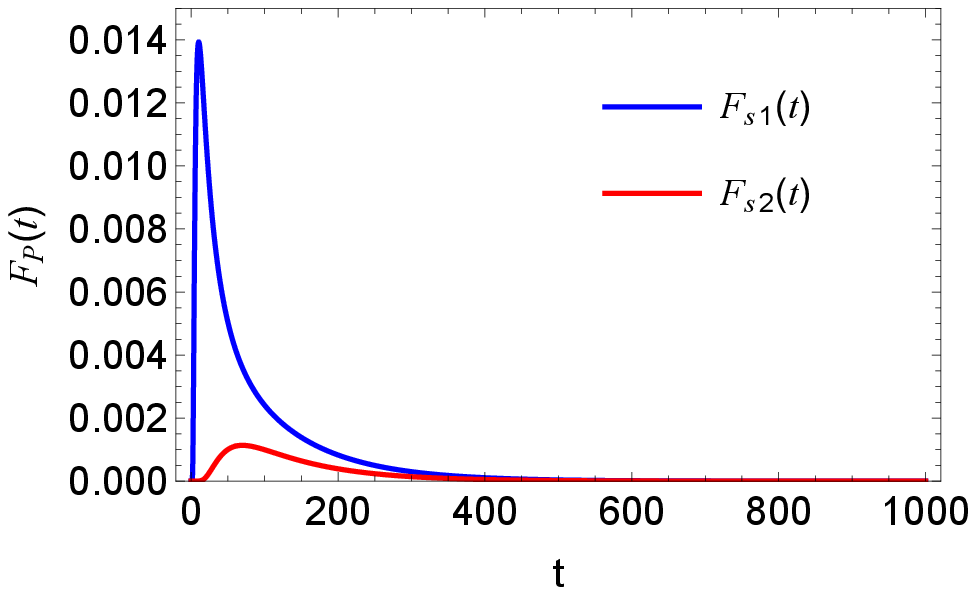}}
\subfigure[]{\label{FP4b}
\includegraphics[width=7cm]{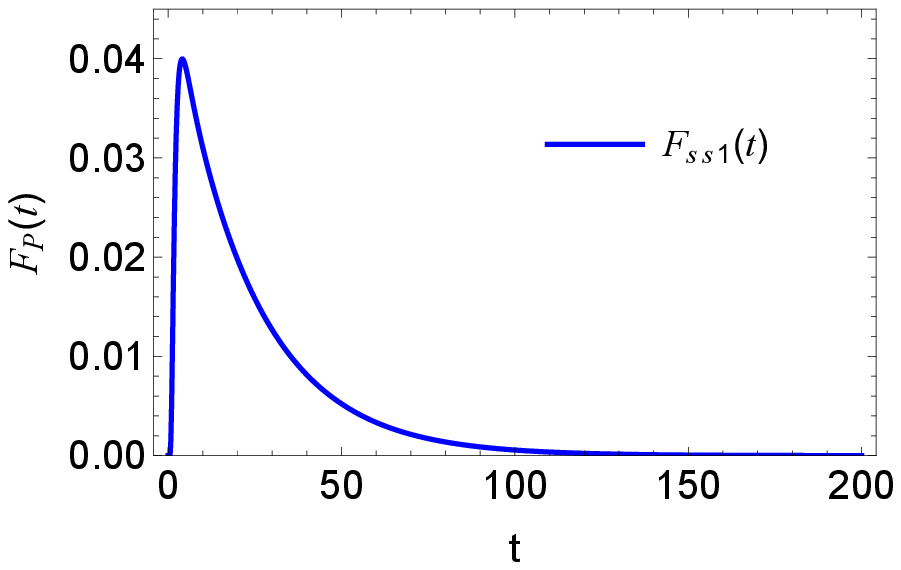}}}
\caption{First passage time distribution at intermediate ensemble temperature $T_{\rm E}=0.00631913$ for initial (a)largest, (b) large, (c) small, (d) smallest black hole respectively.}
\label{FPI}
\end{figure}
\end{center}

\begin{center}
\begin{figure}[!h]
\center{\subfigure[]{\label{LFP4a}
\includegraphics[width=7cm]{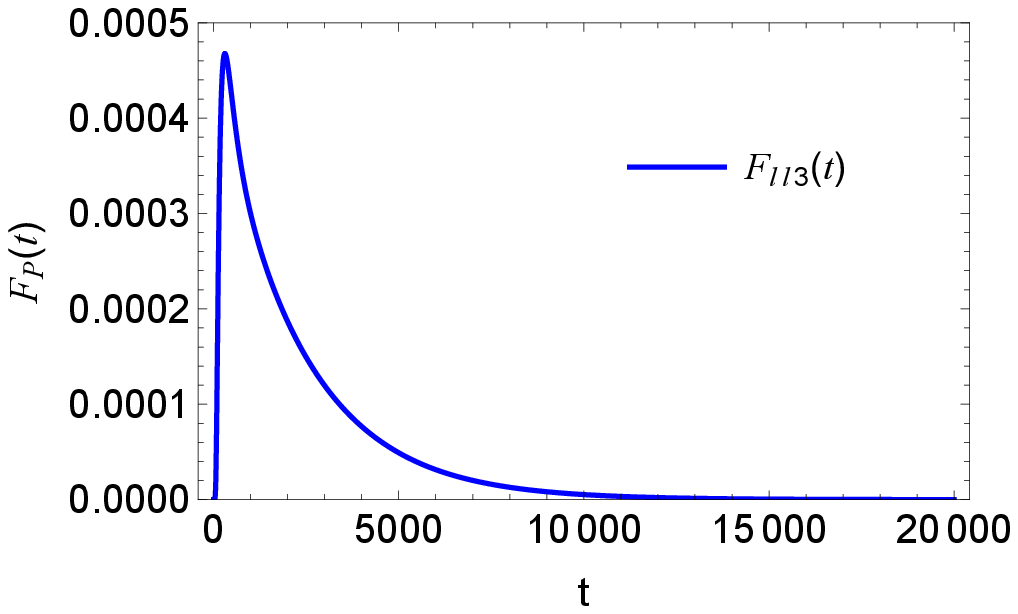}}
\subfigure[]{\label{FP4b}
\includegraphics[width=7cm]{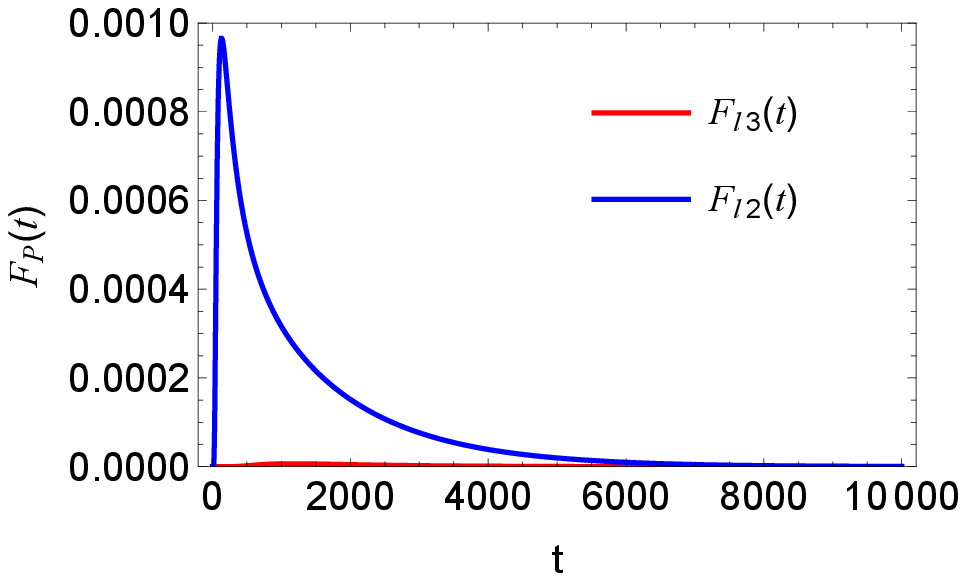}}}
\center{\subfigure[]{\label{FP4a}
\includegraphics[width=7cm]{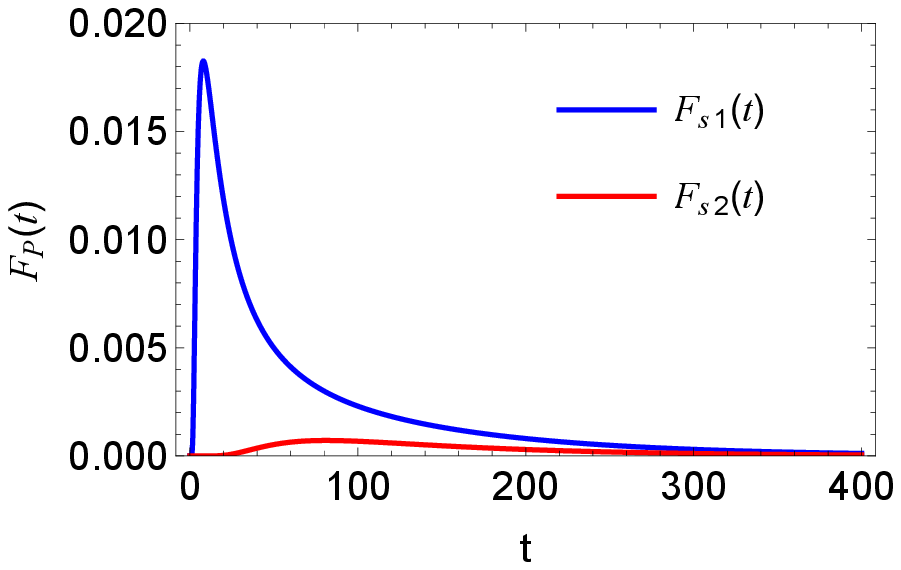}}
\subfigure[]{\label{FP4b}
\includegraphics[width=7cm]{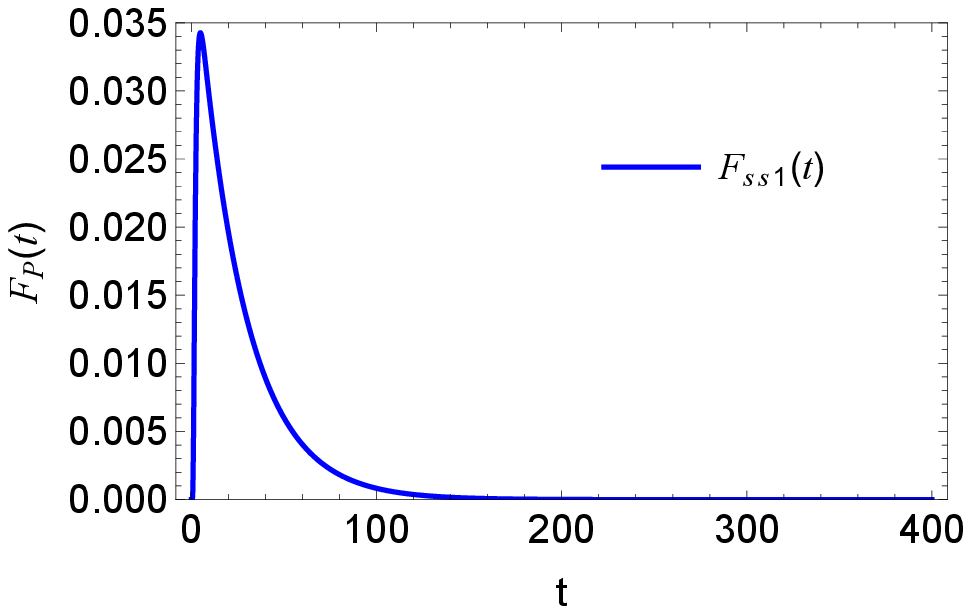}}}
\caption{First passage time distribution at low ensemble temperature $T_{\rm E}=0.006298$ for initial (a)largest, (b) large, (c) small, (d) smallest black hole respectively.}
\label{FPL}
\end{figure}
\end{center}

Let us examine the low ensemble temperature case with $T_{\rm E}=0.006298$, whose Gibbs free energy structure is given by Fig.~\ref{gb}. When starting with the initial black hole state peaked at the largest horizon radius $r_{ll}$, shown in Fig.~\ref{tlowll2a}, the probability is more likely to leak to the state with radius $r_{l}$ at early times; we observe that  $\rho(t, r_{ l})>\rho(t, r_{ s})\gtrsim \rho(t, r_{ ss})$, which is consistent with the relation $t{(ll\to l)}< t{(ll\to s)}\lesssim t{(ll\to ss)}$. However, the situation reverses at late times: from Fig.~\ref{tlowll2a} we observe   that $\rho(t, r_{ ll})<\rho(t, r_{ l})<\rho(t, r_{ s})< \rho(t, r_{ ss})$. This is because the Gibbs free energy has the relation $G_{\rm L}( r_{ ss})<G_{\rm L}( r_{ s})<G_{\rm L}( r_{ l})<G_{\rm L}( r_{ ll})$ at low ensemble temperature, and the final state probability distributions $   \rho(r)\propto e^{-\beta G_{\rm L}(r)}$, which means states with lower free energy have a higher probability of being occupied. The different orderings of the probability distributions for the states with the four stable horizon radii in the early and late stages result in the intersection points of these curves in Fig.~\ref{tlowll2a}, which we call strong oscillatory behaviour. For the  large initial black hole state with radius $r_{l}$, Fig.~\ref{tlowl2b} shows that $\rho(t, r_{ s})\gtrsim \rho(t, r_{ ss})\gg\rho(t, r_{ ll})$ at early times, owing to the phase transition time scale relationship $t{(l\to s)}\lesssim t(l\to ss)<t(l\to ll)$.
The probability distribution function $\rho(t, r)$ for the initial small black hole in Fig.~\ref{tlows2c} exhibits the fastest growth rate at $r_{ ss}$ and the slowest growth rate at $r_{ ll}$ during the early stages of evolution. Specifically, we observe that $\rho(t, r_{ ss}) > \rho(t, r_{ l}) > \rho(t, r_{ ll})$, which can be attributed to the ordering $t({s\to ss}) < t(s\to l) < t(s\to ll)$
of the   transition time scales. 
Similarly, for the initial state with the smallest black hole in Fig.~\ref{tlowss2d}, we find that $\rho(t, r_{ s})$ grows the fastest at early times, followed by $\rho(t, r_{ l})$ and $\rho(t, r_{ ll})$, consistent with the order of time scales for transitions: $t(ss\to s) < t(ss\to l) < t({ss\to ll})$. And for the final states in Fig.~\ref{tlowl2b},  Fig.~\ref{tlows2c}, and  Fig.~\ref{tlowss2d}, we observe the same phenomenon as Fig.~\ref{tlowll2a}, $\rho(t, r_{ ll})<\rho(t, r_{ l})<\rho(t, r_{ s})< \rho(t, r_{ ss})$, which is determined only by the structure of the Gibbs free energy in Fig.~\ref{gb} and is independent of the initial state. Similarly, at high ensemble temperature $T_{\rm E}=0.0063422$, the dynamic behaviours at early times shown in Fig.~\ref{high2d} can also be explained well utilizing the first passage time presented in Table~\ref{ta3}, while the late time behaviour is governed by the off-shell Gibbs free energy, shown in Fig.~\ref{gc}.

\begin{table}[]
\begin{tabular}{|c|c|c|c|c|c|c|}
\hline
  & $H_{ll3}$    & $H_{l3 }$     & $H_{l2 }$     & $H_{s2 }$     & $H_{s1 }$     & $H_{ss1}$     \\ \hline
$T_L$ & 0.0024 & 0.027   & 0.0036  & 0.0082  & 0.0011  & 0.0018  \\ \hline
 $T_I$& 0.015  & 0.015   & 0.0066 & 0.0066 & 0.0016 & 0.0016 \\ \hline
$T_H$ & 0.035 & 0.0055 & 0.011  & 0.0050 & 0.0021 & 0.0013 \\ \hline
  & $W_{ll3}$    & $W_{l3 }$     & $W_{l2 }$     & $W_{s2 }$     & $W_{s1 }$     & $W_{ss1}$     \\ \hline
$T_L$ & 3.3   & 7.7    & 2.2    & 1.9    & 0.54   & 0.40   \\ \hline
$T_I$ & 6.1   & 5.8    & 2.8    & 1.7   & 0.62   & 0.37   \\ \hline
$T_H$ & 8.3   & 3.9    & 3.5    & 1.5   & 0.69   & 0.34   \\ \hline
\end{tabular}
\caption{ Barrier height $H$ and barrier
width $W$ from initial largest ($r=r_{ll}$),  large ($r=r_{l}$), small ($r=r_{s}$) and smallest  ($r=r_{ss}$) black hole states to the adjacent top of
 the barrier $r=r_{m1}, r=r_{m2}$,  and $ r=r_{m3} $. }
 \label{ta1}
\end{table}

\begin{table}[]
\begin{tabular}{|c|c|c|c|c|c|c|}
\hline
  & $t_{ll3}$ & $t_{l3}$ & $t_{l2}$ & $t_{s2}$ & $t_{s1}$ & $t_{ss1}$ 
     \\ \hline
$T_L$&307 &1160 &132& 81  & 8  & 5 
\\ \hline
$T_I$ & 1593 & 912 & 224 & 70  & 10  & 4
\\ \hline
$T_H$ & 3309 & 466 & 363& 58  & 13  & 3
     \\ \hline
\end{tabular}
\caption{ First passage time peak from initial largest ($r=r_{ll}$),  large ($r=r_{l}$), small ($r=r_{s}$) and smallest  ($r=r_{ss}$) black hole states to the adjacent top of
 the barrier $r=r_{m1}, r=r_{m2}$,  and $ r=r_{m3} $. }
 \label{ta2}
\end{table}

\begin{table}[]
\begin{tabular}{|c|c|c|c|}
\hline
  & $t(ll\to l)$ & $t(ll\to s) $& $t(ll\to ss)$ \\ \hline
 $T_L$ & 307    & 439    & 447  \\ \hline
 $T_I$ & 1593   & 1817   & 1827   \\\hline
  $T_H$ & 3309   & 3672   & 3685 \\\hline
  &$ t(l\to s )$ &$ t(l\to ss)$ & $t(l\to ll) $ \\ \hline
 $T_L$& 132    & 140    & 1160    \\ \hline
$T_I$ & 224    & 234    & 912     \\ \hline
$T_H$   & 363    & 376    & 466     \\ \hline
  &$ t(s\to ss)$ &$ t(s\to l )$ &$ t(s\to ll)$ 
\\ \hline
$T_L$ & 8      & 81     & 1241 \\\hline
$T_I$ & 10     & 70     & 982  \\\hline
$T_H$ & 13     & 58     & 524   \\\hline
& $t(ss\to s) $& $t(ss\to l) $& $t(ss\to ll)$ \\ \hline
 $T_L$  & 5      & 86     & 1246    \\ \hline
  $T_I$  & 4      & 74     & 986     \\ \hline
$T_H$  & 3      & 61     & 527     \\ \hline
\end{tabular}
\caption{ First passage time scale from initial largest ($r=r_{ll}$),  large ($r=r_{l}$), small ($r=r_{s}$) and smallest  ($r=r_{ss}$) black hole states to other stable states. }
 \label{ta3}
\end{table}

\begin{center}
\begin{figure}[!h]
\center{\subfigure[]{\label{HFP4a}
\includegraphics[width=7cm]{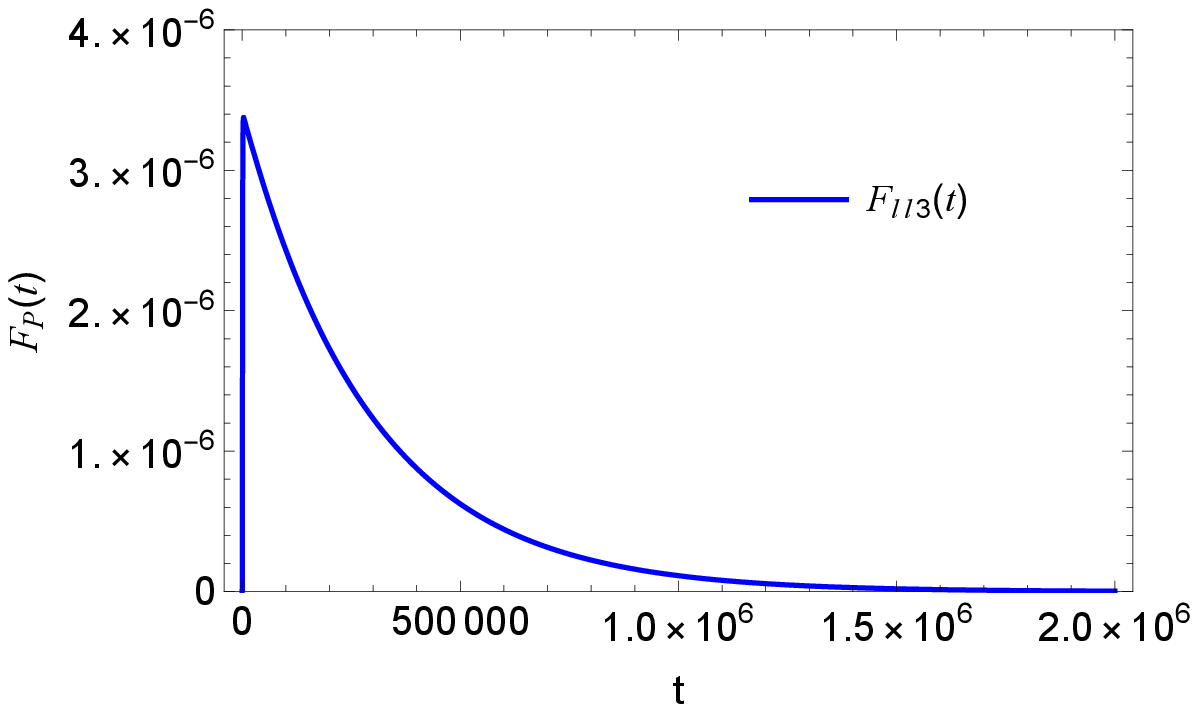}}
\subfigure[]{\label{FP4b}
\includegraphics[width=7cm]{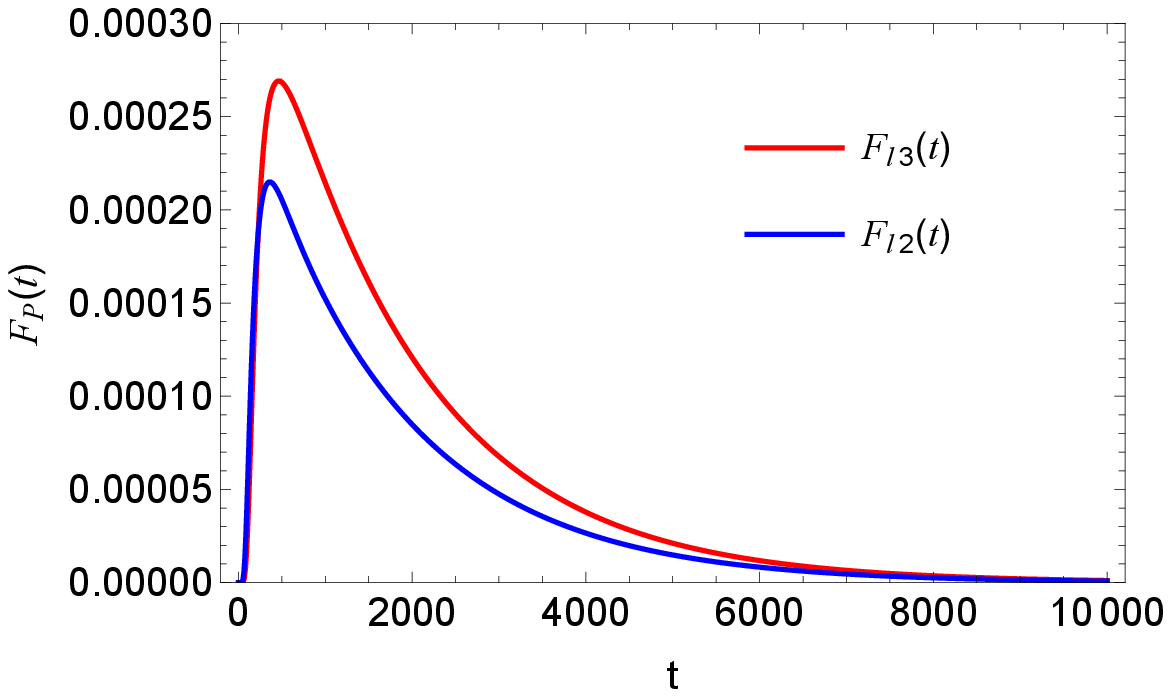}}}
\center{\subfigure[]{\label{FP4a}
\includegraphics[width=7cm]{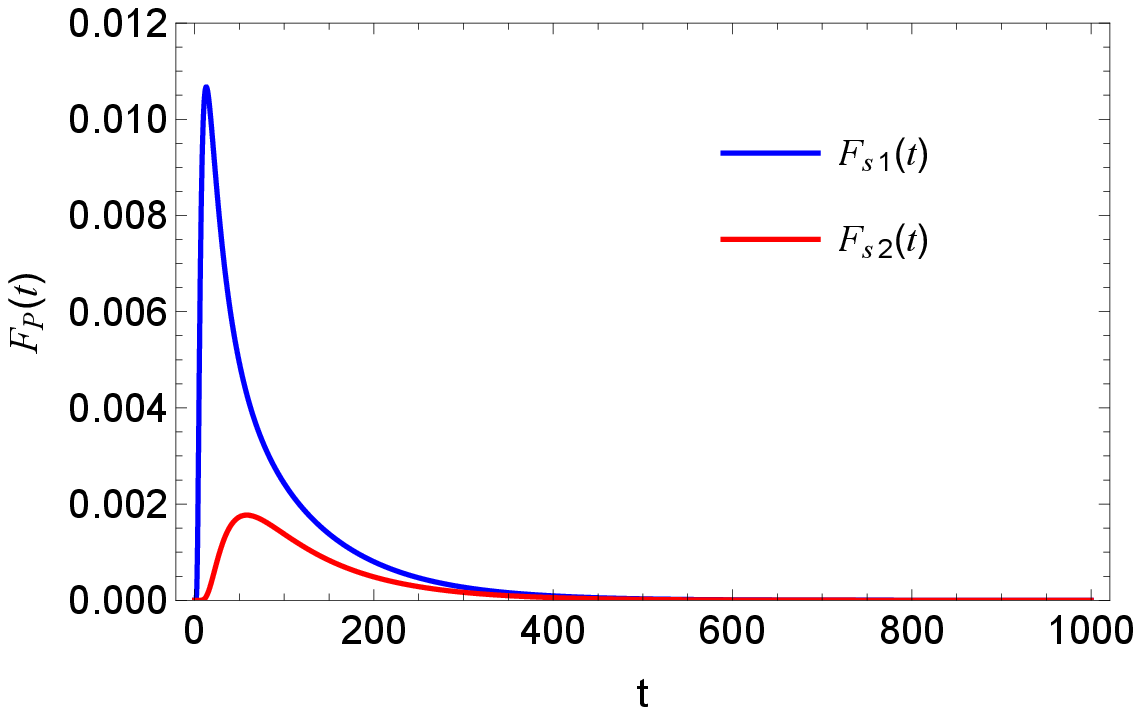}}
\subfigure[]{\label{FP4b}
\includegraphics[width=7cm]{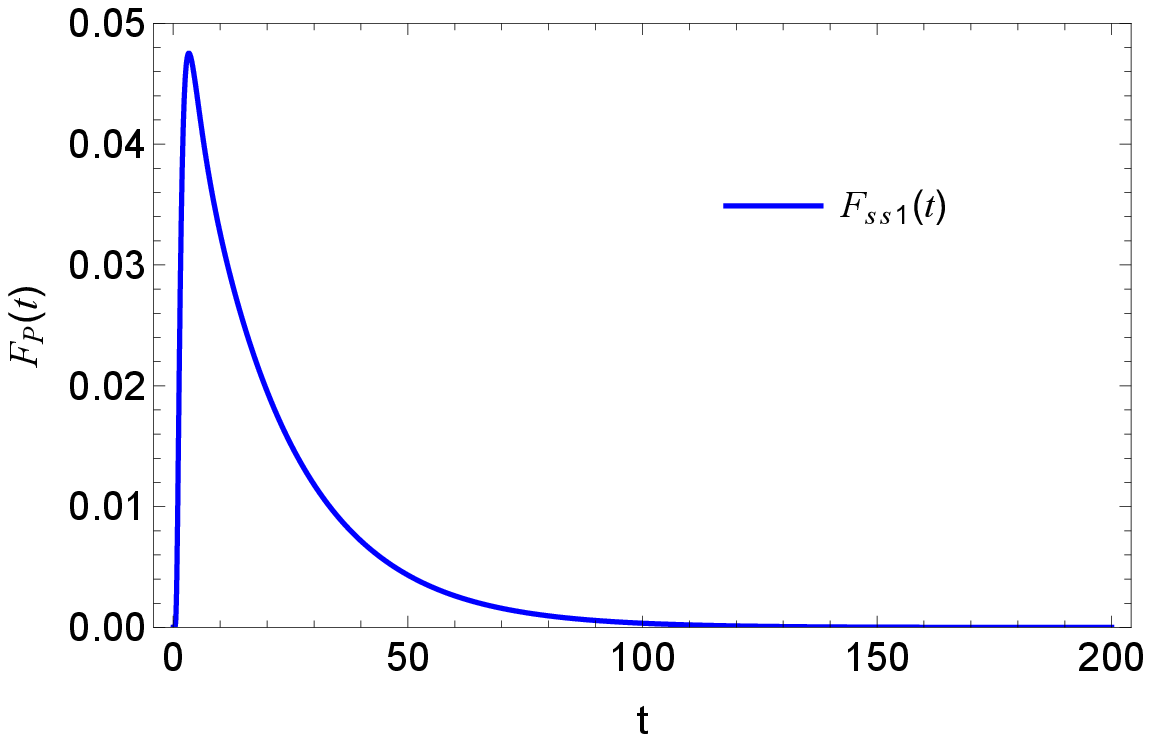}}}
\caption{First passage time distribution at high ensemble temperature $T_{\rm E}=0.0063422$ for initial (a)largest, (b) large, (c) small, (d) smallest black hole respectively.}
\label{FPH}
\end{figure}
\end{center}

We now study the impact of ensemble temperatures on first passage events. Our findings, presented in Table~\ref{ta1} and Table~\ref{ta2}, indicate that as the ensemble temperature rises, the barrier widths for $W_{ll3}, W_{l2}, W_{s1}$ increase, resulting in an increase for the respective first passage times $t_{ll3}, t_{l2}, t_{s1}$. Conversely, higher temperatures cause a decrease in $W_{l3}, W_{s2}, W_{ss1}$, leading to a decline in $t_{l3}, t_{s2}, t_{ss1}$. This provides further evidence that there is a positive correlation between the peak time of first passage time distribution and barrier width. Specifically, we see from Fig.~\ref{Gibbs} that a change in temperature does not substantively change the size of the barrier
between the smallest and small phases,  whereas 
the barrier between the large and largest phases undergoes a huge change.  Consider then the cases where the initial state is peaked at the largest black hole states at different ensemble temperatures. The first passage times to reach the state with radius $r_{ l}$ from  $r_{ ll}$ are denoted by $F_{ll3}$. As shown in Fig.~\ref{Gibbs}, when the ensemble temperature increases, both barrier height and barrier width increase for the initial largest black hole. Comparing Fig.~\ref{LFP4a}, Fig.~\ref{IFP4a}, and Fig.~\ref{HFP4a}, we conclude that as temperature increases, the peak of the first passage time distributions moves rightward and becomes smaller, indicative of the increasing difficulty to surmount higher and wider barriers.  Computing the first passage times, we have $t_{ll3}$=307, 1593, 3309 for low, intermediate, and high ensemble temperature $T_{\rm E}$, respectively. Notably, a significantly long time is needed to reach other states from the largest black hole state at high temperatures. The drastically low value of $F_{ll3}$ and long first passage time explain the final predominance of the largest black hole state at high temperatures. 

\end{widetext}

\section{Conclusions}

Studying black hole thermodynamics and phase transitions  offers valuable insights into the microscopic structure of black holes. We have carried out the first investigation of the dynamical behaviour of a multicritical phase transition, concentrating specifically on    black hole quadruple points in Einstein gravity coupled to non-linear electrodynamics. The quadruple point is characterized by three pairs of local
maxima and minima on a $T$-$r_{+}$ phase diagram whose inflection points are all at the same temperature $T$, corresponding to the temperature at multicriticality.
 On the free energy landscape, with $G_{\rm L}$ as potential and $r_{+}$ as the order parameter, there are four potential wells for quadruple points, representing four coexisting black hole phases.


To study the multicritical phase transition dynamics, we set the initial state to be one of the four stable black hole phases, employing the Smoluchowski equation to solve for the evolution of the probability distribution. We find that the probability density of the initial state diffuses to the other three states, with the diffusion rate depending on the first passage time distribution. 
We find that this distribution in turn is governed by the size of the potential barriers between the states in the off-shell Gibbs free energy. The probability distribution stabilizes at late times, and the probabilities of the four phases are determined by the values of the off-shell Gibbs free energy.

The first passage time distribution has a characteristic pattern where it rapidly rises to a peak within a short time and then quickly declines, indicating that there is a high probability of the first passage event occurring within a short time period. The shorter the corresponding first passage time, the greater the probability density of the initial state diffusing to an adjacent phase.

 The different orderings of the $\rho(t, r)$ of the four phases at early and late times result in  strong oscillatory phenomena:   for example (Fig.~\ref{tlowll2a}), 
  at early times,   $\rho(t, r_{ ll})>\rho(t, r_{ l})>\rho(t, r_{ s})\gtrsim \rho(t, r_{ ss})$, whereas at late times the situation reverses, and  $\rho(t, r_{ ll})<\rho(t, r_{ l})<\rho(t, r_{ s})< \rho(t, r_{ ss})$. The different orderings of the probability distributions for the states with the four stable horizon radii in the early and late stages result in the intersection points of these curves in Fig.~\ref{tlowll2a}, which is  strong oscillatory behaviour.

 Additionally, even with the same orderings, there may exist weak oscillatory phenomena. And orderings of the $\rho(t, r)$ are affected by $G_{\rm L}$. Thus, the evolution of black hole states at the quadruple point is actually determined by the structure of the off-shell Gibbs free energy diagram. We have found that temperature is an important factor that can alter the structure of the off-shell Gibbs free energy diagram, and therefore it also affects black hole phase transitions. 
 
Our results extend previous work carried out for black hole  triple point phase transitions \cite{Wei:2021bwy}. We have concentrated on investigating the dynamic behaviours of the quadruple point under varying ensemble temperatures, showing how these influence the probability distributions 
in a multiphase scenario.
  By analyzing the initial diffusion and  final stable behaviours of the four stable states, we explain the oscillatory behaviours observed during the evolution process.

In conclusion, we investigate intriguing phenomena of black hole phase transitions from the perspective of stochastic dynamics. This study may help to understand the underlying physics of black hole phase transitions and the microscopic structure of black holes. Additionally, the free energy landscape model can also be used to examine the dynamic behaviours of the quintuple points or higher multicritical points \cite{Tavakoli:2022kmo,Wu:2022bdk,Wu:2022plw}, where more complex behaviour may become manifest. \\

{\emph{Acknowledgements}.}---We would like to thank Jerry Wu, Si-Jiang Yang, Shao-Wen Wei, Yu-Xiao Liu, Yong-Qiang Wang, and Niayesh Afshordi for helpful discussions. This work was supported in part by the Natural Sciences and Engineering Research Council of Canada. J.Y. is grateful for support from a
 Mitacs Globalink Graduate Fellowship and encouragements from  Hai-jun Wang and Han-qing Liu.

\end{document}